\documentclass[twocolumn]{aastex63}
\usepackage{CJK}
\usepackage{amsmath}

\def\siggas{$\Sigma_{\rm{gas}}$}
\def\sigSFR{$\Sigma_{\rm{SFR}}$}

\def\etaEh{$\eta_{E,h}$}
\def\etamh{$\eta_{m,h}$}
\def\etaZh{$\eta_{{Z,h}}$}

\def\msun{$M_\odot$}

\def\sigSFR{$\dot{\Sigma}_{\rm{SF}}$}
\def\rmax{$R_{\rm{max}}$}
\def\rout{$R_{\rm{out}}$}
\def\vout{$\widetilde{v}_{\rm{h,t}}$}
\def\msunyr{$M_\odot\ \rm{yr}^{-1}$}

\submitjournal{ApJ}

\shorttitle{SNe-Driven Outflows \& CGM}
\shortauthors{Li et al.}

\graphicspath{{./}{figures/}}

\begin{document}
\begin{CJK*}{UTF8}{gbsn}

\title{How Do Supernovae Impact the Circumgalactic Medium? \\
I. Large-Scale Fountains Around a Milky Way-Like Galaxy }

\correspondingauthor{Miao Li}
\email{mli@flatironinstitute.org}

\author[0000-0003-0773-582X]{Miao Li (李邈)}
\affiliation{Center for Computational Astrophysics, Flatiron Institute, New York, NY 10010, USA}

\author[0000-0002-8710-9206]{Stephanie Tonnesen}
\affiliation{Center for Computational Astrophysics, Flatiron Institute, New York, NY 10010, USA}

\begin{abstract}
{Feedback is indispensable in galaxy formation. However, lacking resolutions, cosmological simulations often use ad hoc feedback parameters. Conversely, small-box simulations, while better resolving the feedback, cannot capture gas evolution beyond the simulation domain. We aim to bridge the gap by implementing small-box results of supernovae-driven outflows into dark matter halo-scale simulations and studying their impact on large scales. Galactic outflows are multiphase, but small-box simulations show that the hot phase (T$\approx$ 10$^{6-7}$ K) carries the majority of energy and metals. We implement hot outflows in idealized simulations of the Milky Way halo, and examine how they impact the circumgalactic medium (CGM). In this paper, we discuss the case when the star formation surface density is low and therefore the emerging hot outflows are gravitationally bound by the halo. We find that outflows form a large-scale, metal-enriched atmosphere with fountain motions. As hot gas accumulates, the inner atmosphere becomes ``saturated'. Cool gas condenses, with a rate balancing the injection of the hot outflows. This balance leads to a universal density profile of the hot atmosphere, independent of mass outflow rate. The atmosphere has a radially-decreasing temperature, naturally producing the observed X-ray luminosity and column densities of O VI, O VII, O VIII. The self-regulated atmosphere has a baryon and a metal mass of $(0.5-1.2)\times 10^{10}M_\odot$ and $(0.6-1.4)\times 10^8 M_\odot$, respectively, small compared to the ``missing" baryons and metals from the halo. We conjecture that the missing materials reside at even larger radii, ejected by more powerful outflows in the past.
 }

\end{abstract}


\section{Introduction}
\label{intro}

The circumgalactic medium (CGM) is the battle field between cosmic accretion and galactic feedback. It is closely related to galaxy formation and the baryon distribution in the universe, where multiple unsolved problems exist. For example, (i) galaxies only contain a small fraction of cosmic baryons \citep{fukugita98,bell03,guo10,moster10,planck18}. Why is galaxy formation so inefficient? and (ii) where are the baryons that are not in galaxies \citep{cen99,bregman07,shull12}? (iii) Galaxies only retain a small fraction of heavy elements (``metals") they have produced \citep[e.g.][]{tremonti04,erb06,andrews13,peeples14,sanchez19}. Where are the metals? (iv) Galaxies have a bimodal distribution, i.e., star-forming galaxies, and massive quenched systems \citep[e.g.][]{kauffmann03,blanton03}. How do galaxies stop forming stars? The dynamical, thermal, and chemical states of the CGM contain vital information about the cosmic baryon cycle and are inextricable from galaxy formation \citep{cen11,shen13,suresh15}. The mass, energy, and metal contents of the CGM are critical for understanding the cosmic baryon distribution \citep[e.g.][and references therein]{tumlinson17,bregman18}.

The CGM has been a central target for many recent observing programs. It has multiphase components, with cool gas around $10^4$ K, warm-hot phase around $10^5$ K, and hot phase ($> 10^6$ K) \citep[e.g.][]{anderson10,chen10,steidel10,tripp11,kacprzak12,bouche12,li13,zhu13,werk14,johnson15,bielby19}.  
CGM properties correlate with galaxy properties. For example, OVI has a very high detection rate in star-forming galaxies, but almost no detection in quiescent galaxies \citep{tumlinson11}.
X-ray luminosities of galaxy coronae scale with star formation rates (SFRs) in galaxies \citep{mineo12,wang16}. The motions of the cool CGM increase with the SF intensity in the galaxy
\citep{lan18,schroetter19,martin19,rudie19}. 
For the Milky Way, a relatively quiescent galaxy with a SFR of 1-3 \msun yr$^{-1}$, the hot and warm-hot phase is seen through O VIII, O VII, and OVI absorption lines along many lines of sight \citep{gupta12,fang15,das19,sembach03}; cool CGM phases are prevalent, with a fraction falling toward the galaxy as high-velocity clouds \citep[e.g.][and references therein]{putman12}. Recent theoretical works have provided clues on the working mechanisms of the CGM, through analytic modeling \citep{maller04,voit17,faerman17,lochhaas18,keller19,stern19,faerman20} and numerical simulations \citep[e.g.][]{shen13,hummels13,fielding17,corlies18,hafen19}.

Energetic feedback from the galaxies plays an essential role in cosmic baryon cycles and galaxy formation. In particular, supermassive black holes and supernovae (SNe) dominate the energy output and can substantially change the CGM, as indicated by recent simulations \citep[e.g.][]{oppenheimer20,zhang20,angles-alcazar17,nelson19}. Our focus in this paper is on SNe. SNe produce copious energy and the majority of metals. They maintain a multiphase ISM \citep{mckee77,cox05}, regulate star formation (SF) \citep[][and references therein]{mckee07}, and drive galactic outflows \citep{maclow99,strickland09,creasey13,li17a,kim18,hu19,fielding18,emerick19}. The outflows transport energy, mass, and metals into the CGM and even into intergalactic space \citep[e.g.][]{songaila96,schaye03}. It is thus important to better understand how the CGM evolves under SNe-driven outflows. Questions of particular interest include: how far can outflows travel? How far can metals go? What fraction of mass/metals are retained in the CGM, break into the intergalactic medium (IGM), or fall back to galaxies?

On the other hand, how SNe feedback works is still an unsolved problem. 
The energy and metal production of SNe are reasonably well-known \citep[][and references therein]{weiler88,arnett95}. Yet, it is not known how SN remnants interact with a multiphase ISM \citep{li15,kim15b,martizzi16,zhang19}, and how much of their energy is used in regulating SF, driving outflows, or is simply dissipated away. Cosmological simulations, due to their general lack of resolution, often use \textit{ad hoc} models for SNe feedback. While the feedback is tuned so that major galaxy properties match the observations, their CGM have very different properties and are sensitive to the feedback models used \citep[e.g.][]{shen13,suresh15,stewart17,davies20}. We believe that a better understanding of SNe feedback and applying a more physically-based feedback model in cosmological simulations are critical next steps to a predictive theory of galaxy formation.

Recently, several groups have simulated how SNe drive galactic outflows from the ISM \citep[e.g.][]{creasey13,li17a,girichidis16b,kim18,fielding18,hu19}. These simulations cover a kpc-scale domain with pc-scale resolution, which is generally able to resolve the cooling radius of SNe-driven blast waves. This is essential for convergence on the properties of the multiphase ISM and outflows, especially the hot phase \citep{simpson15,kim15,hu19}. The outflows, like the ISM, typically have three phases \citep{mckee77}: hot ($10^{6-7}$ K), cool ($\sim 10^{4}$ K), and cold ($\lesssim$ 100 K). Outflow rates of mass, energy, and metals are quantified from small-box simulations. The hot outflows have a larger volume fraction and are faster, whereas the cooler phases occupy a smaller volume and are slower. Quantitatively, however, the outflow fluxes can be different depending on detailed physics, in particular, where SNe explode -- whether in dense clouds or diffuse medium, the spatial-temporal correlation of the SNe \citep{gatto15,martizzi16,li17a,fielding18}, the presence of cosmic rays\citep[e.g.][]{simpson16,girichidis18}, etc. These complexities appear to pose challenges for any simple model of SNe-driven outflows. 

However, simple and consistent results seem to emerge when different phases of outflows are counted separately. \cite{li20} compiled recent results from multiple small-box simulations of SNe-driven outflows. They use different numerical codes and the physics included also vary, such as self-gravity and star formation, other stellar feedback, etc\footnote{These simulations do not include cosmic rays, which, according to recent studies, can change the ISM structure significantly\citep[e.g.][]{girichidis18,farber18}.}. Intriguingly, however, \cite{li20} find consensus that the majority of outflow energy is carried by the hot phase. The energy flux of hot outflows is 2-20 times greater than the cool phases. The cooler phases \textit{usually} dominate the mass fluxes. Seemingly, cool outflows carry mass outside the galaxies. However, the specific energy, i.e., energy per unit mass, of the cool outflows is small. Most cool outflows seen in small-box simulations cannot escape DM halos of $M\gtrsim 10^{9-10}$ \msun, whereas hot outflows can escape halos up to $10^{13}$ \msun\ \citep{li20}. The former is confirmed by recent observation \citep{schroetter19}. For a MW-mass galaxy, most cool outflows will travel only a few kpc above the disk before falling back due to gravity \citep{li17a,kim18}. In contrast, the hot outflows have a specific energy 10-1000 times larger than the cool phases \cite[][and references therein]{li20}, and can therefore travel much farther away from the galaxy. Hot outflows should thus have a much greater impact than the cool outflows for halos with $M\gtrsim 10^{10}$ \msun. Furthermore, hot outflows also carry a great amount of metals \citep{creasey15,li17a,hu19}, which can be transported to large distances. Consequently, it is necessary to study how hot outflows impact the thermal, dynamical and chemical states of the CGM. Besides the power of hot outflows, the fluxes of mass, energy, and metal mass of hot outflows have tight correlations, essentially reducing the three parameters into one. These findings make it promising to implement hot outflows into large-scale simulations. 
 
To see the impact of outflows on larger scales, one needs a much larger volume and a much longer time scale than what small-box simulations can cover. Because of computational expense, this means sacrificing resolution and thus the ability to resolve individual SN remnants. In other words, the launching of the outflows cannot be captured simultaneously while covering their domain of impact (though it is now possible for small galaxies for a short duration \cite[e.g.][]{schneider18,hu19,emerick19}.

Our approach is to take the small-box results, i.e., the averaged outflow rates of energy, mass, and metals, and add hot outflows as a source term to a large simulation box. We then focus on the evolution of the hot outflows on scales from kpc to hundreds of kpc, and investigate how they impact the CGM over cosmic time.

We organize our paper as follows: in Section 2 we briefly present small-box results that we will use in the global simulations. In Section 3 we introduce the setup of the global simulations. In Section 4 we use a fiducial run to illustrate the basic picture of the evolution of the simulated CGM. In Section 5 we examine how varying input parameters affect the results and discuss the emerging universal density profile. In Section 6 we discuss the observational signatures from the simulated CGM and how they relate to the underlying physics. We discuss the implications of our findings in Section 7 and conclude in Section 8.

\section{Results from small-box simulations}

In this section, we will first introduce the parameterization of the outflows, i.e., the loading factors. Then we will summarize the results from small-box simulations, which will be used in our global CGM simulations. 

To quantify the ability of SNe to launch outflows, three dimensionless loading factors $\eta_m$ (mass loading), $\eta_E$ (energy loading) and $\eta_{\rm{Z}}$ (metal mass loading) are defined, 
\begin{equation}
\dot{\Sigma}_{\rm{m}}   = \eta_m \dot{\Sigma}_{\rm{SF}}, \\
\end{equation}
 
\begin{equation}
\dot{\Sigma}_{\rm{E}}   = \eta_E\ E_{\rm{SN}}  \frac{\dot{\Sigma}_{\rm{SF}}}{m_{*}},\\
\end{equation}

\begin{equation}
\dot{\Sigma}_{\rm{Z}} = \eta_{\rm{Z}}\ m_{\rm{Z,SN}} \frac{\dot{\Sigma}_{\rm{SF}} }{m_{*}},
\end{equation}
where $\dot{\Sigma}_{\rm{m}}$, $\dot{\Sigma}_{\rm{E}}$, $\dot{\Sigma}_{\rm{Z}}$ are the outflow rates per area of mass, energy (thermal + kinetic) and metal mass measured from the simulations; $\dot{\Sigma}_{\rm{SF}}$ is the SF surface density, $m_{*}$ is the mass of stars formed to produce one SN, and $m_{\rm{Z,SN}}$ is the metal mass released per SN. From these definitions, $\eta_E$ is the fraction of SN energy that is carried by the outflows, and $\eta_{\rm{Z}}$ is the fraction of metals released by SN that is carried in the outflows (note that in this definition of $\eta_{\rm{Z}}$, the metal flux does not include metals that are in the ISM before SNe explode). 

We use the results from \cite{li17a}, which include runs over a wide range of \sigSFR. The fiducial runs assume a Kennicutt-Schmidt relation for \siggas\ and \sigSFR.
The loading factors are measured for each run at a height of 1-2.5 kpc from the midplane, i.e., above the ISM and SNe explosion sites, and are temporally-averaged over the last 40\% of the simulation time (160 Myr) when the system reaches semi-steady state. Note that unlike the cool phases, the fluxes of hot outflows change little with the height within the simulation domain, due to their large specific energy compared to the depth of the gravitational potential of small boxes \citep[e.g.][]{kim18}. For the hot outflows, \cite{li17a} found
\begin{equation}
\begin{split}
&\eta_{E,h} = 0.25\pm 0.1, \\
&\eta_{\rm{Z,h}} = 0.5\pm 0.1, \\
&\eta_{m,h} \approx 2.1 \left( \frac{\dot{\Sigma}_{\rm{SF}}}{1.26\times 10^{-4}\ \rm{M_\odot\ yr^{-1}\ kpc^{-2}}} \right)^{-0.4}.
    \end{split}
    \label{eq:load_numbers}
\end{equation}
That is, $\eta_{E,h}$ and $\eta_{\rm{Z,h}}$ are almost constant. This is because the hot phase occupies roughly 50\% of the volume in the ISM, for various \sigSFR. These hot bubbles are ``holes'', where metals and energy can easily vent out and become outflows. The mass loading $\eta_{m,h}$ decreases with increasing \sigSFR. This is likely because when gas surface density is high, the ISM is dominantly in the dense molecular phase \citep[e.g.][]{schruba11}, thus it is hard for the ISM to be loaded to the outflows (like it is harder for winds to blow away rocks than leaves).

A constant $\eta_{E,h}$ and a decreasing $\eta_{m,h}$ for higher \sigSFR\ means that the specific energy of the hot outflows is larger when the SF is more intense. Indeed, this is suggested by X-ray observations \citep{wang16,zhang14} and seen in various small-box works \citep[summarized in Fig. 2 of][]{li20}. 
Quantitatively, we can use the terminal velocity $ \widetilde{v}_{\rm{h,t}}$ to represent the specific energy of hot outflows \citep{chevalier85}, 
\begin{equation}
\begin{split}
\widetilde{v}_{\rm{h,t}} \equiv \left(\frac{ 2\dot{\Sigma}_{\rm{E,h}}  }{  \dot{\Sigma}_{\rm{m,h}}  } \right)^{1/2}
&=  548\ \rm{km\ s^{-1}}  \left( \frac{\eta_E}{0.3} \right)^{1/2} \left(\frac{1.0}{\eta_m}  \right)^{1/2} \\
&=  1225\ \rm{km\ s^{-1}}  \left( \frac{\eta_E}{0.3} \right)^{1/2}\left(\frac{0.2}{\eta_m}  \right)^{1/2}.
\end{split}
\label{eq:v_out}
\end{equation}
The above equation assumes $E_{\rm{SN}} =10^{51}$ erg and $m_{*}= 100\ M_\odot$. 
The first example uses \etamh\ for a low \sigSFR$=6.3\times 10^{-3}$ $M_\odot$ yr$^{-1}$ kpc $^{-2}$, which is the approximate mean \sigSFR\ for the MW, and the second example uses \etamh\ for a larger \sigSFR$=0.15-0.71$ $M_\odot$ yr$^{-1}$ kpc $^{-2}$. The terminal velocity of outflows can be compared to the escape velocity of DM halos $v_{\rm{esc}}$ to evaluate whether hot outflows can potentially leave the halo. 
For the MW halo where $v_{\rm{esc}} \approx$ 600 km s$^{-1}$, $\widetilde{v}_{\rm{h,t}} < v_{\rm{esc}}$ for the smaller \sigSFR\ case
while $\widetilde{v}_{\rm{h,t}} > v_{\rm{esc}}$ for the larger \sigSFR.
So the MW-mass halo is of special interest since changing \sigSFR\ can lead to a transition of escape for hot outflows. 

This paper will focus on the case that has $ \widetilde{v}_{\rm{h,t}}< v_{\rm{esc}} $ for a MW-mass halo. In a companion paper, we discuss the case when $ \widetilde{v}_{\rm{h,t}}> v_{\rm{esc}} $. We find that the above two cases result in fundamentally different CGM.

Note that small-box simulations by other groups have somewhat different loading factors, but intriguingly, the three loading factors differ by a similar factor \citep[see discussions in][]{li20}. This makes the global results easily scalable. We will discuss this point in Section \ref{sec:discuss}.

\section{Global Simulation Setup}
\label{sec:method}

In this Section we describe the setup of our global simulations of CGM, using the small-box results for modeling the launch of hot outflows.

\subsection{Simulation code and cooling}

The hydrodynamic equations are solved by the Eulerian code Enzo \citep{bryan14}, using the finite volume piece-wise parabolic method \citep{colella84}. The fiducial box size is 800 kpc on each side. We use a static refinement throughout the simulation. The spatial resolution is progressively higher toward the center of the box, which is 0.5 kpc for the inner (50 kpc)$^3$, 1.0 kpc for the inner (100 kpc)$^3$, and so on. The dependence of global CGM properties on the resolution is discussed in the Appendix. 

We use the Grackle library to calculate the cooling rate of the gas\footnote{\url{https://grackle.readthedocs.io/}} \citep{smith17}. We assume the extragalactic UV background from \cite{haardt12} at redshift 0 in Grackle when determining the ionization levels from equilibrium calculation. The cooling is metallicity-dependent. Throughout this paper, we adopt a solar metallicity $Z_\odot\equiv$ 0.01295 which is the default value in Grackle.

\subsection{Gravitational potential}

\begin{figure}
\begin{center}
\includegraphics[width=0.45\textwidth]{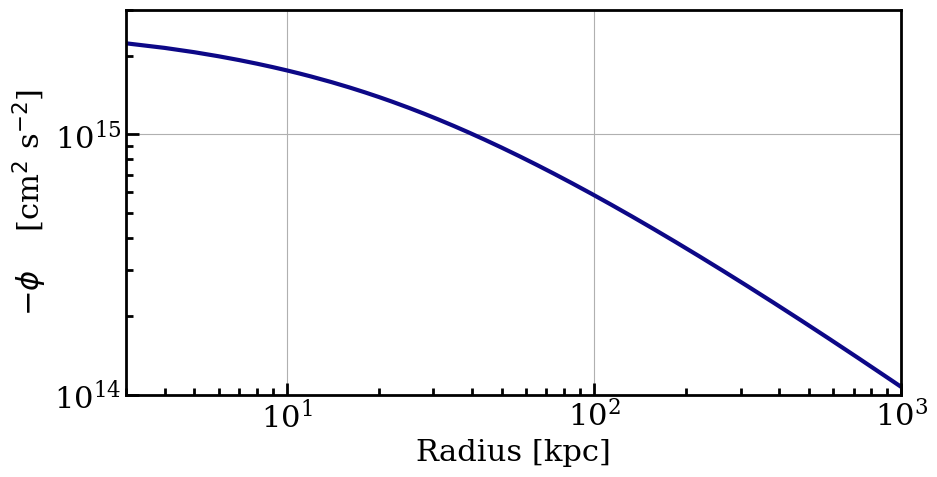}
\includegraphics[width=0.45\textwidth]{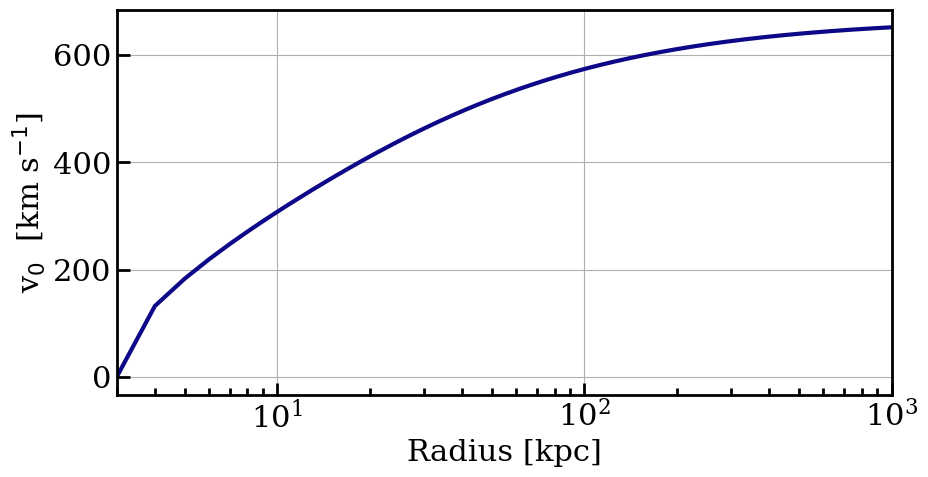}
\caption{Top panel: Gravitational potential of the MW halo in our simulation, $-\phi (R)$. Bottom panel: $v_0(R)\equiv \sqrt{2[\phi(R)-\phi(3\ \rm{kpc})}]$.  The potential includes that from the DM halo, the galactic disk and the stellar bulge. }
\label{f:v_esc}
\end{center}
\end{figure}

The galaxy is located at the center of the box, with the disk in the $x-y$ plane. We use a static gravitational potential. (We simulate the CGM for 8 Gyr, over which the halo mass change by less than a factor of 2 \citep[e.g][]{mcbride09}, so a static halo potential is a reasonable approximation.)  The potential includes a DM halo, a stellar disk, and a stellar bulge. The parameters of the potential follow those of the MW. The DM halo is assumed to have a \cite{burkert95} profile, with a central density of 2.71 $\times 10^{-24}$ g cm$^{-3}$ and a core radius of 10 kpc \citep{nesti13}. The mass distribution of the stellar disk has a Plummer-Kuzmin functional form \citep{miyamoto75}, with a mass of 5.5$\times 10^{10} M_\odot$, a scale radius of 3.5 kpc, and a scale height of 0.7 kpc. The galactic bulge is modeled as a spherical \cite{hernquist93} profile with a mass of $10^{10} M_\odot$ and a scale radius of 1.3 kpc.

The top panel of Fig. \ref{f:v_esc} shows the gravitational potential as a function of radius, $-\phi (R) \equiv \int_0^{R} g(R')dR'$
where $g$ is the gravitational field. The bottom panel shows $v_0(R)\equiv \sqrt{2[\phi(R)-\phi(3\ \rm{kpc})}]$, i.e., the velocity of outflows at the launching site in order to reach a radius $R$ (assuming a ballistic evolution). To escape from the halo (R$\gtrsim$ 250 kpc), the outflow at the galaxy disk needs to have $v_0 \sim$  620 km s$^{-1}$. This is a simple estimate when there is no gas in the halo. With some pre-existing gas, the estimated $v_0$ is a lower limit.

\subsection{Initial conditions}

\begin{figure}
\begin{center}
\includegraphics[width=0.5\textwidth]{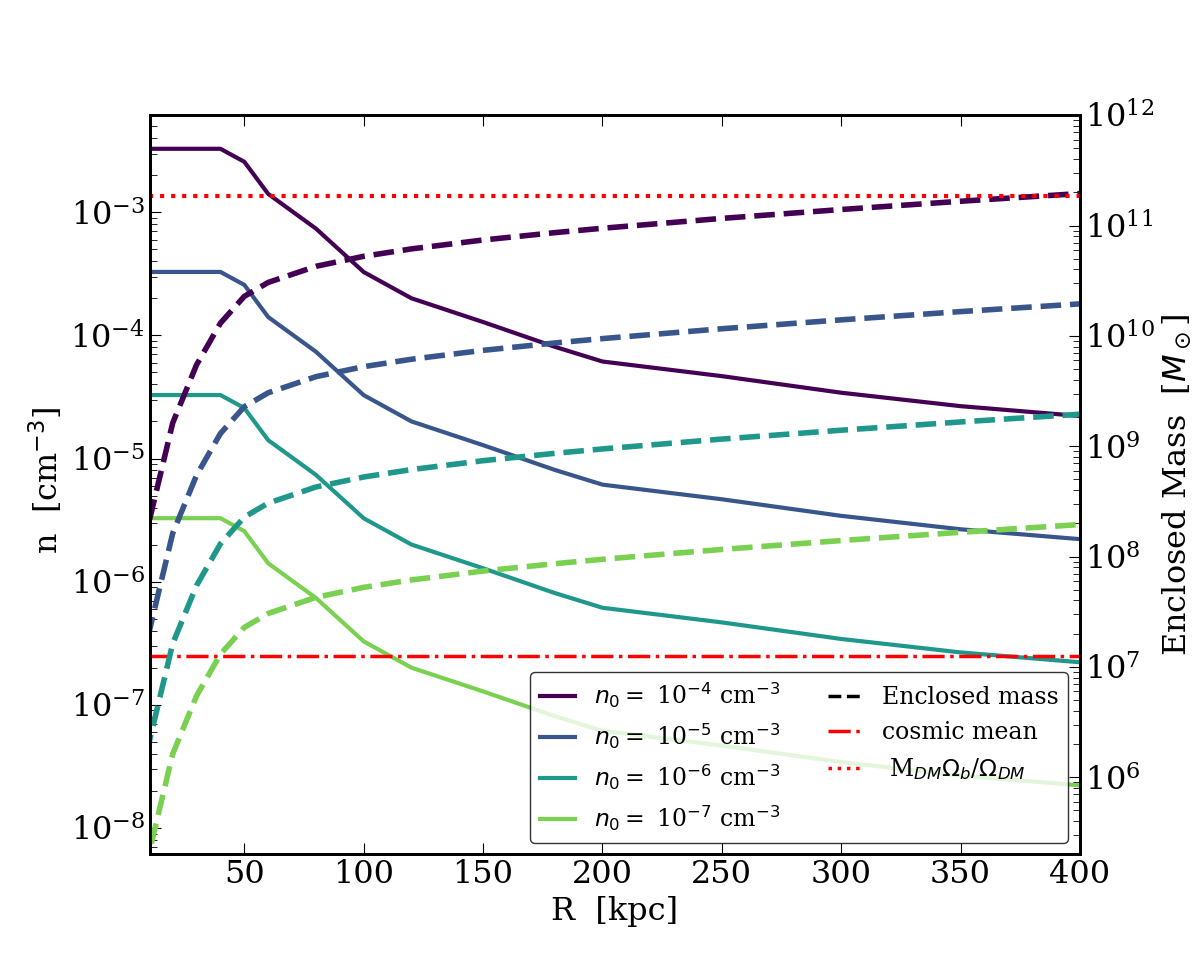}
\caption{Initial condition of radial density profile (solid lines) and enclosed mass (dashed lines). Different colors indicate different $n_0$, the mean density at $R<200$ kpc. The dotted line indicates the baryon mass associated with a DM halo of $10^{12}M_\odot$, assuming the cosmic ratio of dark matter to baryons \citep{planck18}. The dash-dotted line indicates the mean density of the cosmic baryon at redshift 0. }
\label{f:init_condition}
\end{center}
\end{figure}

The initial gas in the simulation box only includes a hot component. We do not include a cool gaseous disk within the galaxy. The gas has a uniform temperature of $10^6$ K, similar to the virial temperature of the MW-mass halo, and a uniform low metallicity $0.2 Z_\odot$. Gas density is set to be in hydrostatic equilibrium with the DM halo potential, with an inner core of uniform density at $R<40$ kpc. This core is artificial, leaving room for a future development of a hot gaseous disk component with rotation; but we have tested 
its exact size does not affect our results (as long as it is small). 
The normalization factor of initial gas density is parameterized by $n_0$, defined as the mean number density of gas within $R=200$ kpc. This is a free parameter in our simulations, and we vary it from $10^{-7}-10^{-4}$ cm$^{-3}$.  
Fig. \ref{f:init_condition} shows the initial density profile as a function of radius (solid lines). The enclosed mass is shown by the dashed lines using the right $y$-axis. The horizontal dash-dotted line indicates the mean density of the cosmic baryons at redshift 0. The dotted line indicates the total mass of baryons associated with a DM mass of $10^{12} M_\odot$, assuming a cosmic baryon fraction \citep{planck18}. Given the uncertainty of the actual initial condition, our range of $n_0$ includes these two limits.

\subsection{Implementation of outflows}

In our simulations, we do not model star formation in the galaxy disk. In fact, as mentioned before, there is no cool gaseous disk to begin with. Instead, we add outflows by hand at a location above/below the disk. 
In terms of where and how often to add outflows, we try to mimic the SF process in a disk galaxy. The basic picture is that SF happens in clusters; each star cluster has a life-span of a few tens of Myr, during which outflows are launched; the location of the star cluster is random within the galaxy disk. Holding this picture in mind, we add the outflows in the following way: the outflows are injected as discrete events to a small region. The locations of outflows are different each time and are randomly selected within the galaxy. The time intervals between these SF events are constant, $\Delta t=$9.9 Myr.

For each SF event, the injected region for outflows is two hemispheres above and below the disk whose coordinates ($x$, $y$, $z$) satisfy:
\begin{equation}
    \begin{cases}
    (x-x_o)^2 + (y-y_o)^2 +(z- z_o)^2 \leqslant R_o^2,\ \rm{for}\ z\geqslant z_0;  \\
    (x-x_o)^2 + (y-y_o)^2 +(z+ z_o)^2 \leqslant R_o^2 ,\ \rm{for}\ z\leqslant -z_0, \\
    \end{cases}
    \label{eq:outflow_location}
\end{equation}
where ($x_o$, $y_o$) is the center of this SF event in the disk, which is  randomly chosen within a circle of $R_{\rm{SF}}$ on the midplane. The hemispheres have a radius of $R_o$, which is set to be 3 kpc. 
Note that the outflows are only added at $|z|\geqslant z_o$. We choose to add the outflows at $z_0=$ 3 kpc above the plane because (1) in small-box simulations, the loading factors are measured at a few kpc above the plane; (2) later when cool gas forms in the CGM, it will fall to the mid-plane with a height of a few cell sizes, below where the outflows are added. This prevents adding the outflows to the dense, cool gas layer, and thus avoids the ``overcooling'' problem commonly seen when applying feedback in coarse-resolution simulations.

As mentioned before, only hot outflows are added. 
The outflow rate of mass, energy, and metal mass are given by the following equations: 
\begin{equation}
    \dot{M}_{\rm{out}} = \eta_m\  \dot{M}_{\rm{SF}},
    \label{eq:M_dot}
\end{equation}

\begin{equation}
    \dot{E}_{\rm{out}} =\eta_E    E_{\rm{SN}} \left(\frac{\dot{M}_{\rm{SF}}}{m_*} \right),
    \label{eq:E_dot}
\end{equation}

\begin{equation}
    \dot{M}_{\rm{Z,out}} = \eta_Z\ m_{\rm Z,SN} \left(\frac{\dot{M}_{\rm{SF}}}{m_*}\right) + \dot{M}_{\rm{out}}  Z_{\rm{ISM}}, 
    \label{eq:Met_dot}
\end{equation}
where $Z_{\rm{ISM}}$ is the metallicity of the ISM that is being ``entrained'' into the outflows, $m_{\rm Z, SN}$ is the metal yield per SN which we assume to be 1.5 $M_\odot$, and $m_*=100$\msun. For each SF event, the amount of mass, energy, and metal mass added as source terms are simply $\dot{M}_{\rm{out}}  \Delta t$, $\dot{E}_{\rm{out}}  \Delta t$, and $\dot{M}_{\rm{Z, out}} \Delta t$, respectively. The injected mass, energy, and metals are uniformly distributed within the two hemispheres. Given energy conservation, there is a choice of the division of outflow energy into the kinetic form and the thermal one. However, we find that the results are insensitive to this choice \footnote{The insensitivity is because the injection region has a much higher energy density than the surroundings, so it will invariably undergo a near-free expansion initially and self-adjust its energy partition along the way. Note that this insensitivity may appear different from traditional galaxy simulations where the form of feedback usually matters. In traditional galaxy simulations, feedback is applied to the galaxy disk with high gas densities. Thus if the resolution is not high enough, thermal deposition will result in the overcooling problem\citep{katz92}, whereas applying kinetic energy instead can partly alleviate the problem \citep[e.g.][]{schaye15,simpson15,pillepich18}. In contrast, here we add outflows in the halo region where gas density is very low ($<10^{-2}$ cm$^{-3}$), thus gas will not suffer from the overcooling issue even when the deposited energy is 100\% thermal. The added outflows will expand and adjust their energy partition in the subsequent evolution, similar to a SN remnant evolving into the Sedov solution.} and we have used a 50-50 partition ($f_k$ $=0.5$, where $f_k$ is the fractional energy in the kinetic form for added outflows). The velocity directions of outflows are along the along the radial direction of the hemispheres and pointing outward. The outflows are added to the pre-existing gas in a momentum-conserving fashion, which we detail below.

In this paper, we use the following loading factors: $\eta_m =$ 1.0, $\eta_E=$0.3, $\eta_Z=$0.5, for an assumed \sigSFR$=6.3\times 10^{-3}$ $M_\odot$ yr$^{-1}$ kpc $^{-2}$ (Eq. \ref{eq:load_numbers}). To be simple, we have adopted a constant $\dot{M}_{\rm{SF}}$ over time. We assume $Z_{\rm{ISM}}$ as a constant in the simulations, 0.8 $Z_\odot$.  Using this $Z_{\rm{ISM}}$ together with Eq. \ref{eq:M_dot} and \ref{eq:Met_dot}, the metallicity of our added outflows is 1.38 $Z_\odot$.
The constant metallicity of outflows is likely an over-simplification, since the ISM is being enriched over cosmic time. But as a first step, we want to keep the model simple, with the plan to increase the model complexity for future work.
The radius of SF region is $R_{\rm{SF}}=$ 8 kpc (note the difference between $R_{\rm{SF}}$ and $R_o$). This indicates that the SF is widespread in the galaxy disk, in contrast to the case where the SF is concentrated at the center, which we will discuss in a companion paper. 

For reproducibility, we summarize how we implement outflows as follows: \\
(i) Every $\Delta t=$ 9.9 Myr, a random position ($x_0$, $y_o$) is chosen within a circle of a radius $R_{\rm SF}$. \\ 
(ii) Source terms of outflows, $\Delta M = \dot{M}_{\rm{out}}  \Delta t$, $\Delta E =\dot{E}_{\rm{out}}  \Delta t $, $\Delta M_Z = \dot{M}_{\rm{Z, out}} \Delta t$ are evaluated through Eqs. (\ref{eq:M_dot})- (\ref{eq:Met_dot}).\\ 
(iii) Given that the total volume of the injection region $V = 4\pi R_o ^3/3$ and the uniform distribution of outflows within that volume, the mass density, kinetic energy density, thermal energy density, and metal density of outflows are thus $\Delta \rho = \Delta M/V$,  $\Delta e_{\rm k}=f_k \Delta E/V$,  $\Delta e_{\rm th}=(1-f_k) \Delta E/V$,   $\Delta \rho_Z = \Delta M_Z/V$, respectively. 
(iv) The momentum flux of outflows is then $\Delta \overrightarrow{p} = \hat{r} ( 2 \Delta e_k \Delta \rho)^{1/2} $.\\
(v) Apply the addition of density, momentum, and metal mass density to each cell within the injection region, as defined by Eq. (\ref{eq:outflow_location}). The quantities prior to/after outflows-adding is indicated without and with a prime. 
\begin{equation}
\begin{split}
    \rho' = \rho + \Delta \rho; \\
     \rho'_Z = \rho_Z + \Delta \rho_Z; \\
     \overrightarrow{p'} =  \overrightarrow{p} + \Delta  \overrightarrow{p}
     \end{split}
     \label{eq:rho_p_add}
\end{equation}
(vi) For each cell, calculate the current kinetic energy density $e_k'$.  Note that there is a loss of kinetic energy during this momentum-adding procedure, that is, $\Delta e_{\rm k,loss} = e_k + \Delta e_k - e_k' >$0. (vii) This loss of kinetic energy is then added as thermal energy. That is, for each cell, apply the thermal energy addition as
\begin{equation}
   e'_{\rm th} =  e_{\rm th} + \Delta e_{\rm th} + \Delta e_{\rm k,loss}
   \label{eq:eth_add}
\end{equation}
In this way, we preserve momentum and total energy when adding outflows. We show in Appendix \ref{sec:flux} that the outflows are added in simulations at the expected rate.

The SFR and $n_0$ for each run are input parameters. We use the nomenclature of ``($n_0$)SFR($\dot{M}_{\rm{SF}}$)" to denote each run. For example,  ``3e-6SFR3" means $n_0 = 3\times 10^{-6}$ cm$^{-3}$ and a SFR of 3 $M_\odot$ yr$^{-1}$.

\section{Results}

In this section we use a fiducial run, 1e-6SFR3, to illustrate the basic picture of CGM evolution with  SNe-driven outflows launched from the disk. The initial density $n_0 =10^{-6} \rm{cm}^{-3}$ is relatively small; therefore the outflow evolution is mostly affected by the gravitational potential. We will discuss the effect of varying $n_0$ and SFR in Section \ref{sec:discuss}. 

First we estimate the distance outflows can travel using a simple energy argument. From Eq. \ref{eq:v_out}, the terminal velocity of the outflows $v_{h,t}$ is 548 km s$^{-1}$ given the loading factors we use. By comparing $v_{h,t}$ to the potential (Fig. \ref{f:v_esc}),
\begin{equation}
   \frac{1}{2} \widetilde{v}_{\rm{h,t}}^2 = \phi(R_{\rm{out}}) - \phi(3\ \rm{kpc}) ,
   \label{eq:rout}
\end{equation}
we get 
\begin{equation}
    R_{\rm{out}}\approx \rm{60\ kpc}.
\end{equation}
This means that from a simple energy argument, outflows injected 3 kpc from the galaxy center should go out to a radius of 60 kpc. Note that if $v_{h,t}$ is larger by 13\%, the hot outflows can escape from the halo.

\subsection{Hot atmosphere}

\begin{figure*}
\begin{center}
\includegraphics[width=1.0\textwidth]{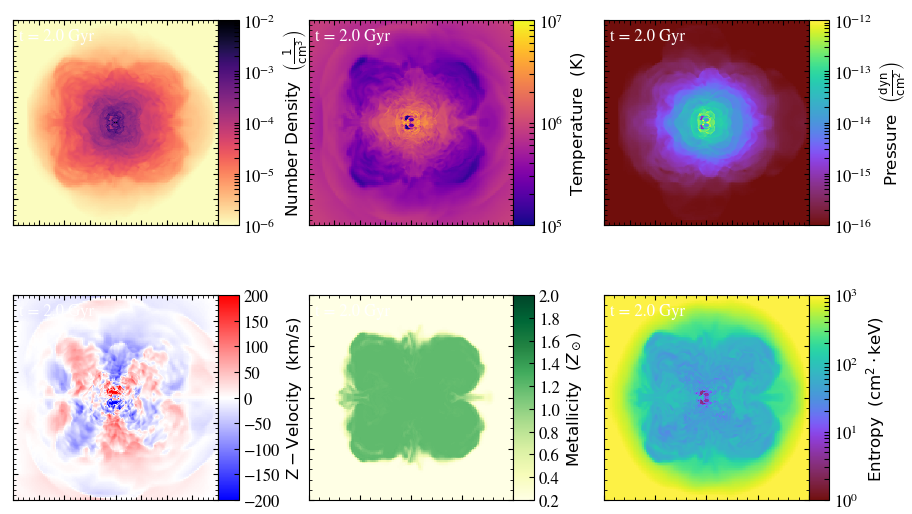}
\caption{Edge-on slices of the fiducial run 1e-6SFR3 at $t=2$ Gyr. The plot is sliced through the center of the box. The horizontal direction is $y$ and the vertical is $z$. The slices are 400 kpc on each side. The hot outflows form a nearly spherical, metal-enriched atmosphere, which has fountain motions. }
\label{f:slice_SFR3}
\end{center}
\end{figure*}

To give a visual impression, Fig. \ref{f:slice_SFR3} shows slices of the fiducial run at 2 Gyr. Each slice is taken at the $y-z$ plane through the center of the box, i.e., the ``galaxy'' is viewed edge-on. The horizontal direction of the slice is $y$ and the vertical is $z$.  The length scale of the slices is 400 kpc on each side (the whole box is twice as large). 
The hot CGM is roughly symmetric around the $z=0$ plane, which is expected due to the symmetry of the initial condition and the outflow implementation. The metallicity slice shows a clear boundary at $R\approx$ 140 kpc, between an enriched inner region and a low-metallicity outer part. This marks the distance that the outflows have reached. Beyond that radius, gas largely remains at the initial condition state, though there are spherical structures seen in temperature, radial velocity, and entropy, which are weak shocks/sound waves that the outflows drive into the pre-existing gas. The slice of $z$-velocity indicates that the metal-enriched region is a large-scale fountain flow, with both positive and negative velocity components within each side of the galaxy. The density, temperature and pressure show a radial decrease within $R\approx 140$ kpc. On the other hand, the entropy is nearly uniform within the enriched region. 
The central region, $R\lesssim$ 10 kpc, has a lower temperature, pressure, and entropy, but high outward velocities. This is because outflows are over-pressured when added and undergo an initial expansion; at around 10 kpc the gas passes through a shock-like jump condition and becomes part of the hot atmosphere.

\begin{figure}
\begin{center}
\includegraphics[width=0.45\textwidth]{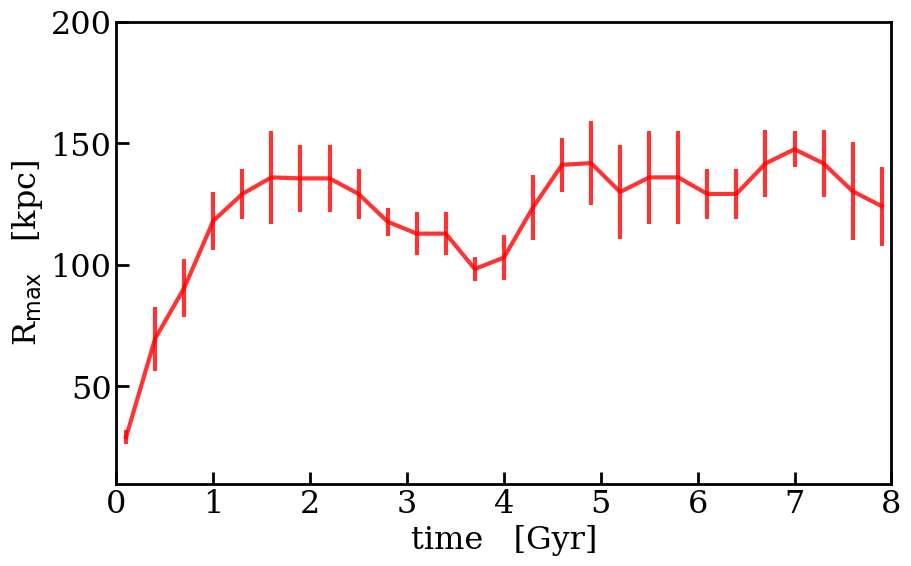}
\caption{Maximum radius metals reach, \rmax, as a function of time for the fiducial run 1e-6-SFR3. The error bars indicate the standard deviation of \rmax\ at different polar angles. \rmax\ first rises to reach about 140 kpc and stays roughly constant after that.
}
\label{f:rmax_t}
\end{center}
\end{figure}

We use \rmax\ to describe the maximum radius of the metal-enriched gas, which is defined as the radius at which the gas metallicity drops to twice that of the pre-existing gas along a given direction. 
Fig. \ref{f:rmax_t} shows how \rmax\ changes over time.
The error bars indicate the standard deviation of \rmax\ at different polar angles, which are small compared to \rmax. This confirms the spherical shape of the outflows. 
The fiducial run shows that \rmax\ rises to about 140 kpc at $t\approx$ 1.5 Gyr, and does not change much after that. There are some fluctuations of \rmax\ over time, which is due to the cooling episode of the hot atmosphere, which we will focus on in the next subsection. The maximum distance the outflows reach, 140 kpc, is larger than the ballistic estimate from Eq. \ref{eq:rout}, which indicates that outflows do not evolve in isolation. We will discuss this more in Section \ref{sec:rmax}. The value of \rmax\ is well converged with respect to the numerical resolution, which we show in the Appendix \ref{sec:res}.

\begin{figure}
\begin{center}
\includegraphics[width=0.5\textwidth]{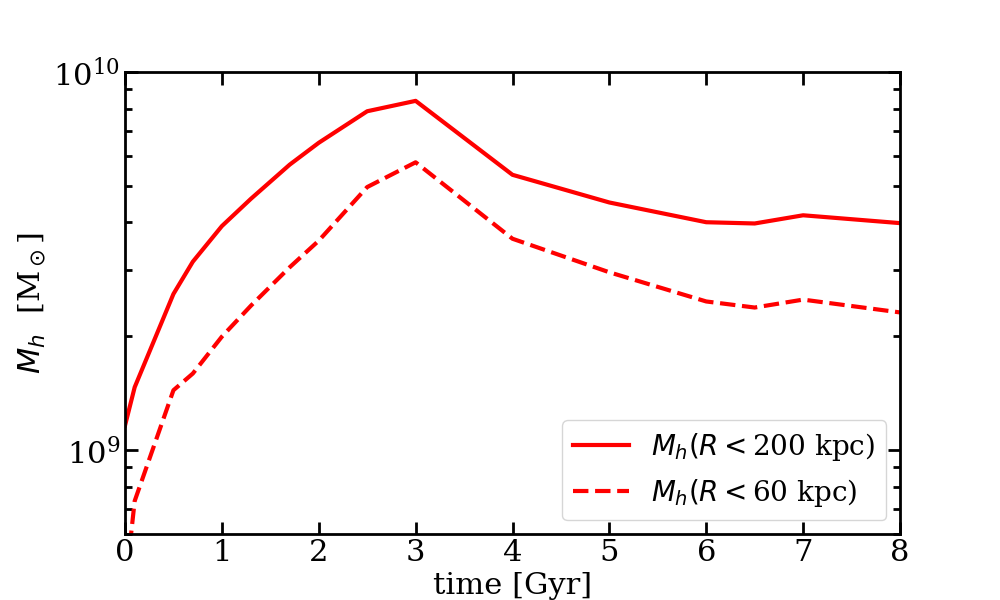}
\includegraphics[width=0.45\textwidth]{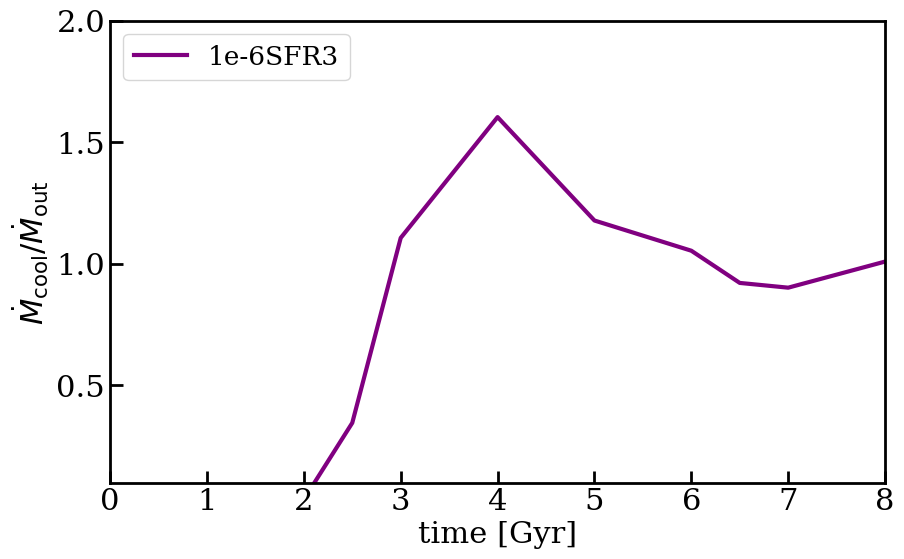}
\caption{Top panel: mass of hot CGM ($T>3\times 10^4$ K) as a function of time for the fiducial run 1e-6SFR3. The solid line indicates the enclosed mass at $R<$ 200 kpc and the dashed line is the enclosed mass at $R<$ \rout\ (where \rout$=$60 kpc). $M_h$ first increases to the maximum, then slightly decreases and settles to a constant value.
Bottom panel: Ratio between the formation rate of cool phase and the hot outflow rate. At $t\gtrsim$ 2.5 Gyr, the ratio is about 1, indicating that the condensation of cool gas condensation balances the injection of hot outflows.
}
\label{f:Mh_Mcool}
\end{center}
\end{figure}

The top panel of Fig. \ref{f:Mh_Mcool} shows the amount of hot CGM as a function of time. Only gas with T $>3\times 10^4$ K is included\footnote{In this paper ``hot gas'' denotes gas with T $>3\times 10^4$ K, which is volume-filling. Gas with $T\lesssim2\times 10^4$ K is defined as ``cool gas".}.
The solid line indicates the enclosed mass within $R<$ 200 kpc and the dashed line is the enclosed mass at $R<$ \rout\ (where \rout$=$60 kpc). 
$M_h (R<200\ \rm{kpc)}$ first increases, reaches the maximum at 8$\times 10^{9}$ M$_\odot$, and then slightly decreases and settles at 4-5$\times 10^{9}$ M$_\odot$. $M_h(R<60$ kpc) is about 60\% of $M_h(R<200$ kpc), meaning that about 40\% of the outflows reach $R>$\rout. The hot outflows are being injected into the box at a constant rate throughout the simulation. Why does the total mass of hot gas in the CGM not keep increasing? It is because as gas accumulates, gas density increases, so does the cooling rate. At some point, the inner hot CGM has a large enough cooling rate that cool phase starts to condense out of the hot atmosphere. That is, the hot atmosphere is ``saturated''.

The bottom panel of Fig. \ref{f:Mh_Mcool} shows the ratio between the formation rate of the cool phase, $\dot{M}_{\rm{cool}}$, and the injection rate of the hot outflows, $\dot{M}_{\rm{out}}$. The cool phase starts to form at around 2 Gyr. The ratio $\dot{M}_{\rm{cool}}/\dot{M}_{\rm{out}}$ quickly rises to around unity within less than 1 Gyr. After that, the ratio stays close to unity. As a result, the amount of hot gas in the CGM remains roughly constant.

\subsection{Cool gas condensation}

\begin{figure}
\begin{center}
\includegraphics[width=0.5\textwidth]{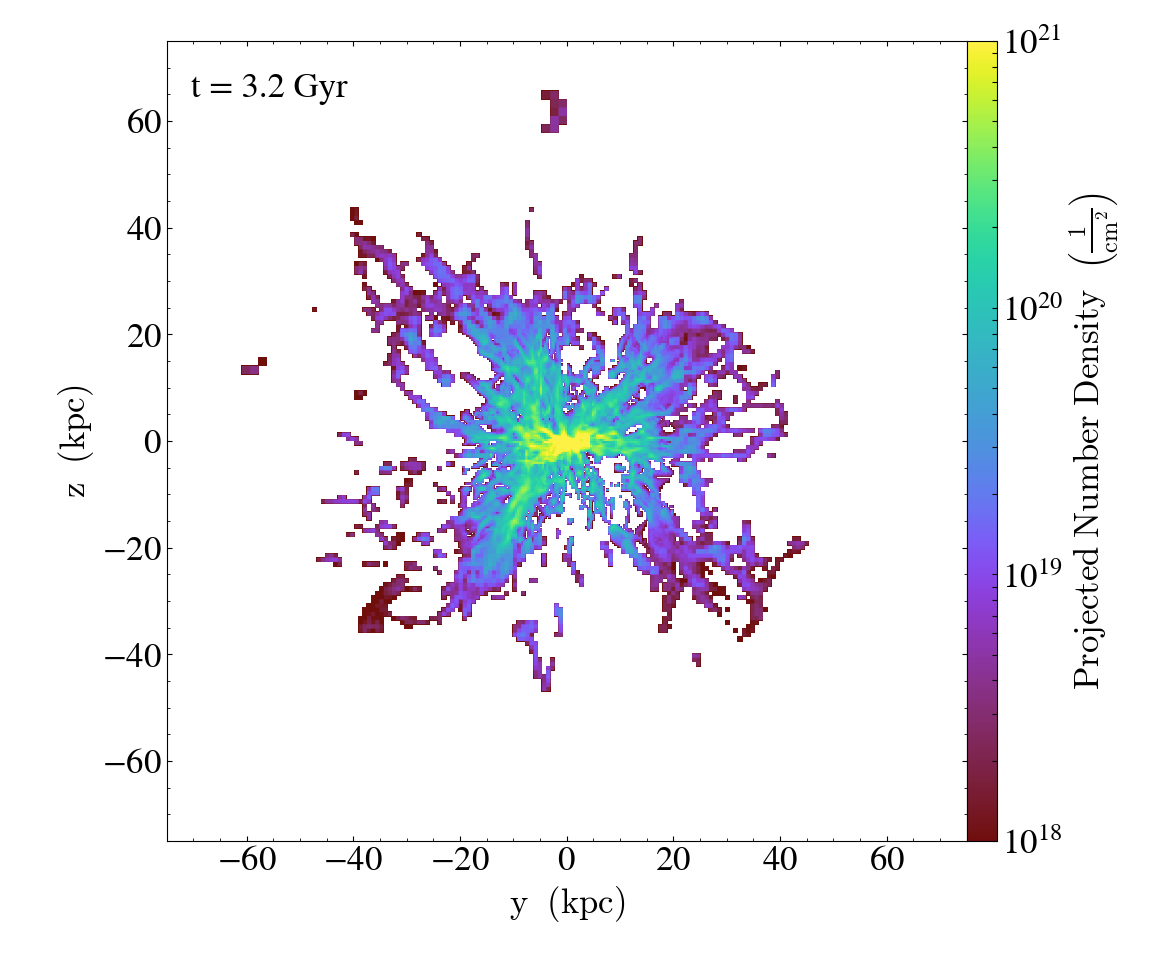}
\caption{Projection of cool gas ($T<3 \times 10^4$K) at 3.2 Gyr for the fiducial run 1e-6SFR3. Galaxy is viewed edge-on. Most of the cool CGM is found at $R\lesssim$ 50 kpc.
}
\label{f:proj_cool}
\end{center}
\end{figure}

Fig. \ref{f:proj_cool} shows the projected density of cool gas ($T<3 \times 10^4$K) at t$=$ 3.2 Gyr. The projection is in the x-direction, i.e., the ``galaxy'' is viewed edge-on. Most of the cool gas is at $R\lesssim$ 50 kpc. In contrast to the smooth hot flows, the cool gas is lumpy and shows filamentary structure. The cool gas has a larger covering fraction at smaller impact parameters.

\begin{figure}
\begin{center}
\includegraphics[width=0.5\textwidth]{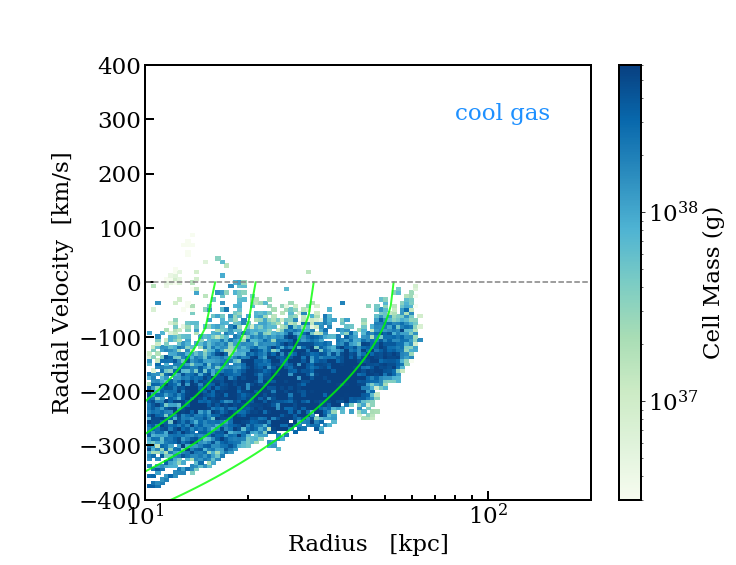}
\includegraphics[width=0.5\textwidth]{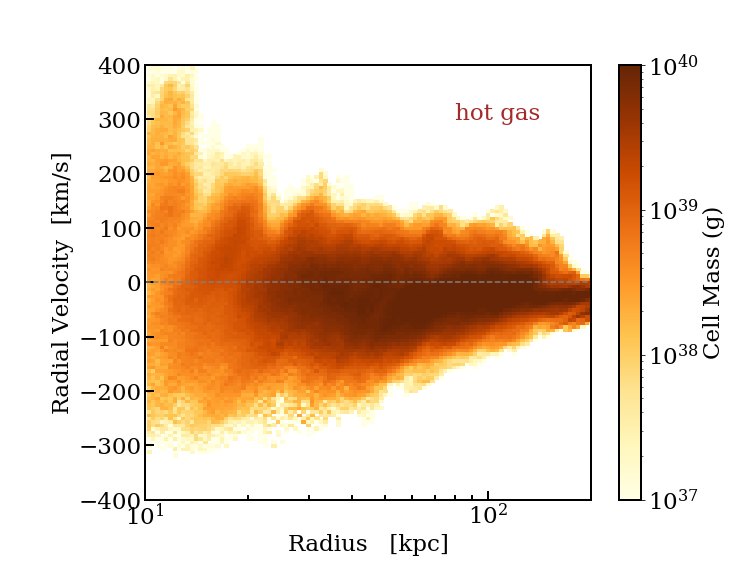}
\caption{Radius versus radial velocity of the cool (top panel) and the hot (bottom panel) gas at $t=$3 Gyr, for the fiducial run 1e-6SFR3. Positive of radial velocity means the gas is moving toward larger radii. The solid lines on the top panel indicate trajectories of free fall at a few radii. While the hot gas moves both outward and inward (fountain), the cool gas mainly falls back ballistically.}
\label{f:vr_r_hot_cool}
\end{center}
\end{figure}

Fig. \ref{f:vr_r_hot_cool} shows the radial velocity versus radius, color-coded by gas mass for the fiducial run. The top panel is for the cool gas and the bottom panel for the hot. The snapshot is taken at $t=3$ Gyr. Outward motions are positive. The hot gas has both outward and inward motions, confirming a fountain flow. In contrast, the cool gas has only negative radial velocities. The inflow velocity increases with decreasing radius. The overplotted solid lines indicate trajectories of free fall beginning at a few radii, given the gravitational potential (Fig. \ref{f:v_esc}). Unlike the hot gas, the cool clumps are not pressure-supported, and fall nearly ballistically toward the center of the potential well. At $R=10$ kpc, the velocity of the cool gas is 100-350 km s$^{-1}$. This is similar to the observed intermediate- or high-velocity clouds in the MW \citep{putman02,mcclure09,lehner10} or infalling gas for external galaxies \citep{zheng17}. 

The lack of a positive-velocity component in cool gas indicates that cool gas does not form when hot outflows are expanding outward. This is different from the picture that \cite{thompson16} and \cite{schneider18} propose about forming cool outflows en route from hot winds. The main difference is that the density (related to mass loading factor) of our outflows is not large enough, thus the cooling time of the hot outflows is much longer than their dynamical time. Instead, the formation of cool gas in our simulations happens significantly later when the hot CGM has accumulated enough mass to be saturated, i.e., when the inner halo has a sufficiently short cooling time of $\sim$Gyr.  

Cool gas dropping out of the hot atmosphere is a natural way to explain the high-velocity clouds in the MW \citep{field65,shapiro76,bregman80,maller04,voit19}. 
If they drop out of an atmosphere formed from SNe-driven hot outflows, they would have the same high metallicity as these outflows, which in our case is $1.4 Z_\odot$. Observationally, some of these high-velocity clouds are highly enriched, with near solar or super-solar metallicity \citep[e.g.][]{zech08}. This highly-enriched cool phase should originate from the galaxy, since accreted materials would have low metallicities. From small-box simulations, the cool outflows launched from the disk typically cannot go very far, $<$ 10 kpc \citep{li17a,kim18}, and the fall-back velocity is limited by the launching velocity which is a few tens of km s$^{-1}$. In contrast, cooling of the hot CGM can happen farther away from the galaxy, and naturally has larger fall-back velocities. Based on our simulation, we argued that observed highly-enriched high-velocity clouds are thus very likely from the cooling of the highly-enriched hot CGM.

\subsection{Summary of basic picture}

From the discussion in this section, we see that gravitationally-bound hot outflows form a hot atmosphere in the DM halo, with large-scale fountain motions. The hot atmosphere is highly-enriched, with the same metallicity as the hot outflows. The atmosphere is nearly spherical. As hot gas accumulates in the halo, the atmosphere becomes ``saturated'', i.e., radiative cooling becomes important. Cool gas then forms and falls back to the galaxy. We stress that this picture holds only when the outflows are gravitationally bound. As we will show in a separate paper, when the outflows are gravitationally unbound, regions occupied by outflows are generally not spherically symmetric, but bipolar, and the resultant hot CGM does not have the same metallicity as the hot outflows due to their mixing with low-metallicity pre-existing halo gas. The mixing with pre-existing gas is not a significant effect for hot outflows in this paper, because their entropy is lower than that of the pre-existing gas. As a result of little mixing, there is a clear division between the inner atmosphere formed almost completely from the hot outflows, and an outer layer of pre-existing gas. This division is clearly seen from the metallicity slice in Fig. \ref{f:slice_SFR3}.

\section{Effects of Varying parameters}

In this section, we discuss how CGM properties change when we vary input parameters, specifically $n_0$ and SFR, while the loading factors are the same. We discuss the fountain height, formation of cool gas, and density profile of the hot CGM. 

\subsection{Fountain height \rmax }
\label{sec:rmax}

\begin{figure}
\begin{center}
\includegraphics[width=0.5\textwidth]{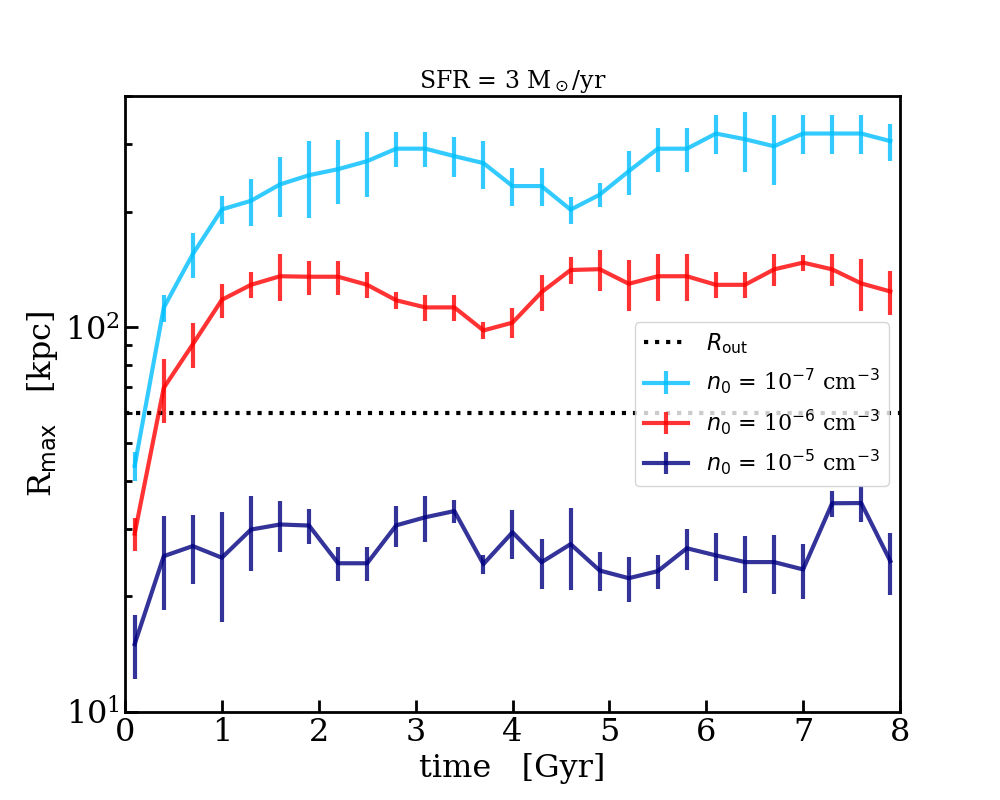}
\includegraphics[width=0.5\textwidth]{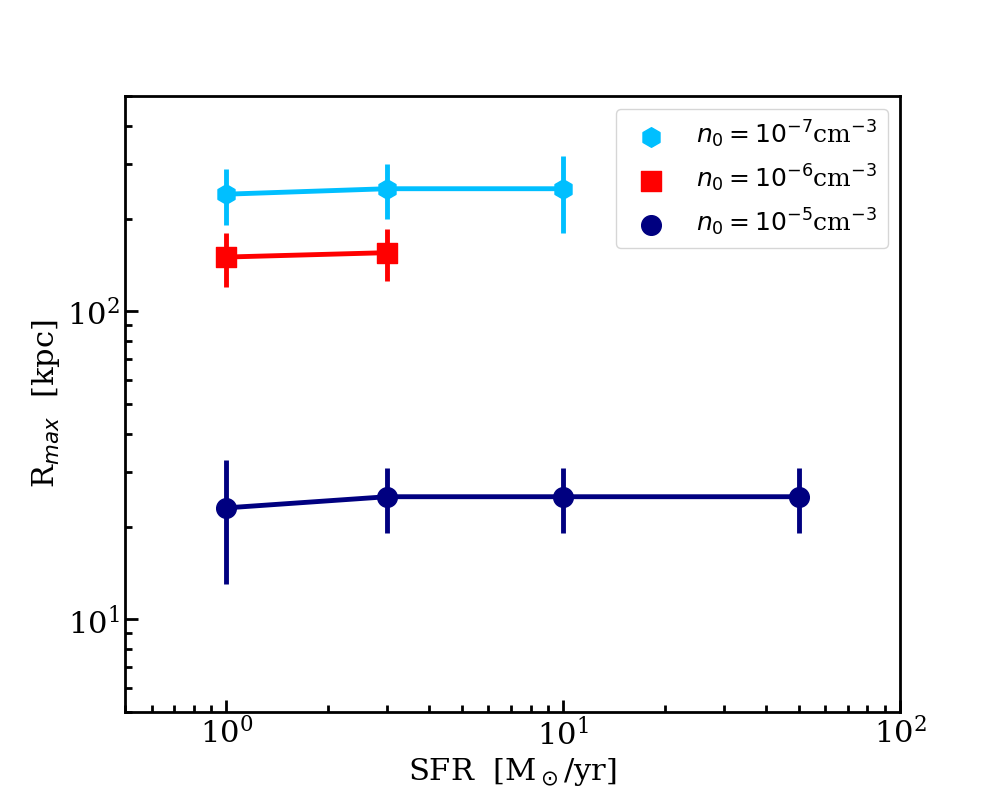}
\caption{Top panel: time evolution of \rmax\ for runs with different pre-existing gas density $n_0$. The error bars indicate variations at different polar angles. The dashed line indicate \rout. \rmax\ decreases with $n_0$.
Bottom panel: \rmax\ (after reaching plateau) as a function of SFR for various $n_0$. The final \rmax\ does not depend on SFR. The error bars indicate the variations over time and polar angles. 
}
\label{f:rmax_diff_n0_SFR}
\end{center}
\end{figure}

As discussed above, \rmax\ quantifies how far outflows can reach and spread metals. The top panel of Fig. \ref{f:rmax_diff_n0_SFR} shows how \rmax\ changes with time for different $n_0$. The SFR for all three runs is 3 \msunyr. Like the fiducial case, \rmax\ reaches a constant value after a dynamical time. The final \rmax\ is smaller when $n_0$ is larger. This is expected since larges $n_0$ means a heavier weight of pre-existing gas. For a sufficiently small $n_0 \lesssim 10^{-6}$ cm$^{-3}$, \rmax$>$\rout, that is, outflows reach further than the simple energy argument dictates (see Equation \ref{eq:rout}). This implies that each parcel of outflow does not evolves in isolation, but energy transfers from small radii to large, through, for example, (weak) shocks\footnote{In fact, if we do not have a sustaining outflow injection, but only allow a few SF events, then \rmax$\sim$ \rout.}. But note that even though \rmax\ can be larger than 200 kpc, i.e., outflows occupy a very large volume, the mass outside \rout\ is limited. For example, Fig. \ref{f:Mh_Mcool} shows that only 40\% of the outflow mass is beyond \rout.

When $n_0$ is large, $\gtrsim 10^{-5}$ cm$^{-3}$, \rmax$<$\rout. This means that the pre-existing gas is sufficiently heavy that outflows are confined by its weight. The thermal pressure of pre-existing gas also plays a role in confining outflows, but one can prove that this is minor compared to its weight.

The bottom panel of Fig. \ref{f:rmax_diff_n0_SFR} shows the final \rmax\ (values after reaching plateau) as a function of SFR for different $n_0$. The error bars indicate the standard deviation of \rmax\ over time and polar angles. For a given $n_0$, the final \rmax\ does not depend on SFR. This may seem counter-intuitive: the outflow rates of energy and mass scale linearly with SFR, so why do larger fluxes of outflows not have a stronger effect? This is because of radiative cooling. The cooling time is short compared to the dynamical time over which \rmax\ increases. Note that with the loading factors unchanged, the runs with high SFRs, 10 or 50 \msunyr, probe an unrealistic parameter space, but are important for understanding the evolution of outflows in the halo. In reality, such a high SFR generally means a much larger \sigSFR, thus the loading factors would be different.

\subsection{Cool phase formation \& hot CGM mass}

\begin{figure}
\begin{center}
\includegraphics[width=0.5\textwidth]{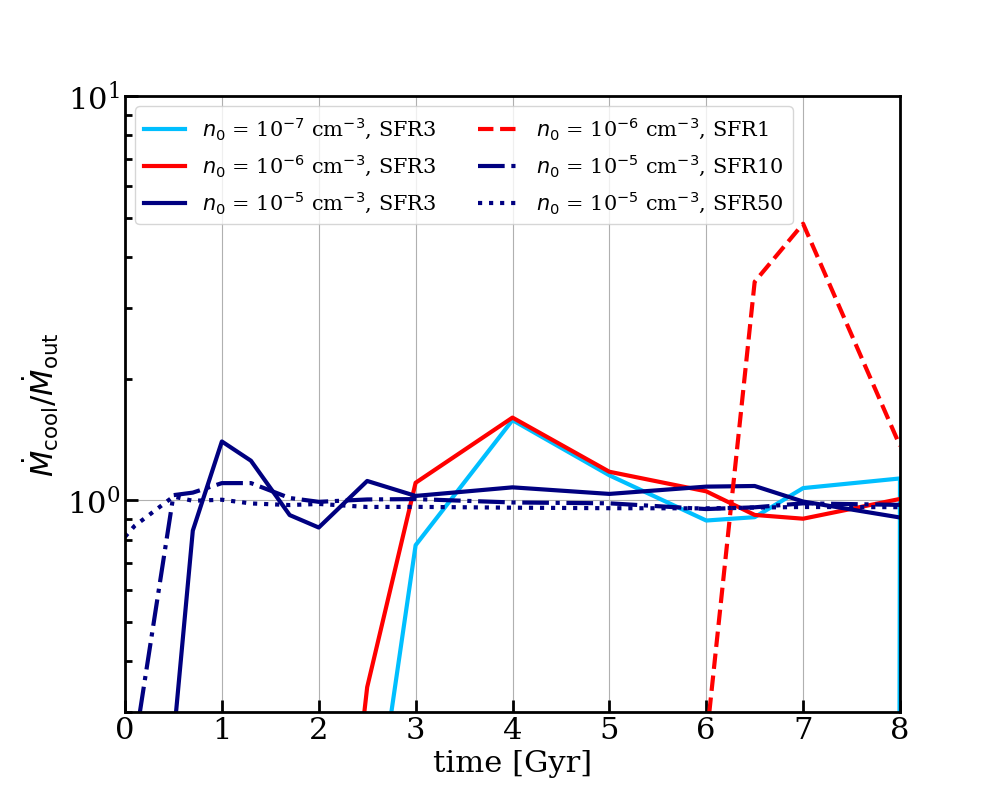}
\caption{Ratio between the formation rate of the cool gas  $\dot{M}_{\rm{cool}}$ and the hot outflow injection rate $\dot{M}_{\rm{out}}$ as a function of time for various SFR and $n_0$. Once the cool gas starts to form, the ratio is about unity. }
\label{f:Mdot_cool_t}
\end{center}
\end{figure}

As in the bottom panel of Fig. \ref{f:Mh_Mcool}, Fig. \ref{f:Mdot_cool_t} shows $\dot{M}_{\rm{cool}}/\dot{M}_{\rm{out}}$ as a function of time for runs with various $n_0$ and SFR. 
Once the cool phase starts to form, its formation rate is roughly equal to that of hot outflow injection rate. This is true for all $n_0$ and SFR we have tried. For a given $n_0$, runs with lower SFRs have later onsets of cool phase formation. This is understandable since it takes longer for the hot CGM to saturate when the mass injection rate is small. 
For the same SFR, the larger $n_0$ is, the earlier the cooling starts. This is because less outflow mass is needed for the saturation of the CGM when there is already some mass in the halo.

\begin{figure}
\begin{center}
\includegraphics[width=0.5\textwidth]{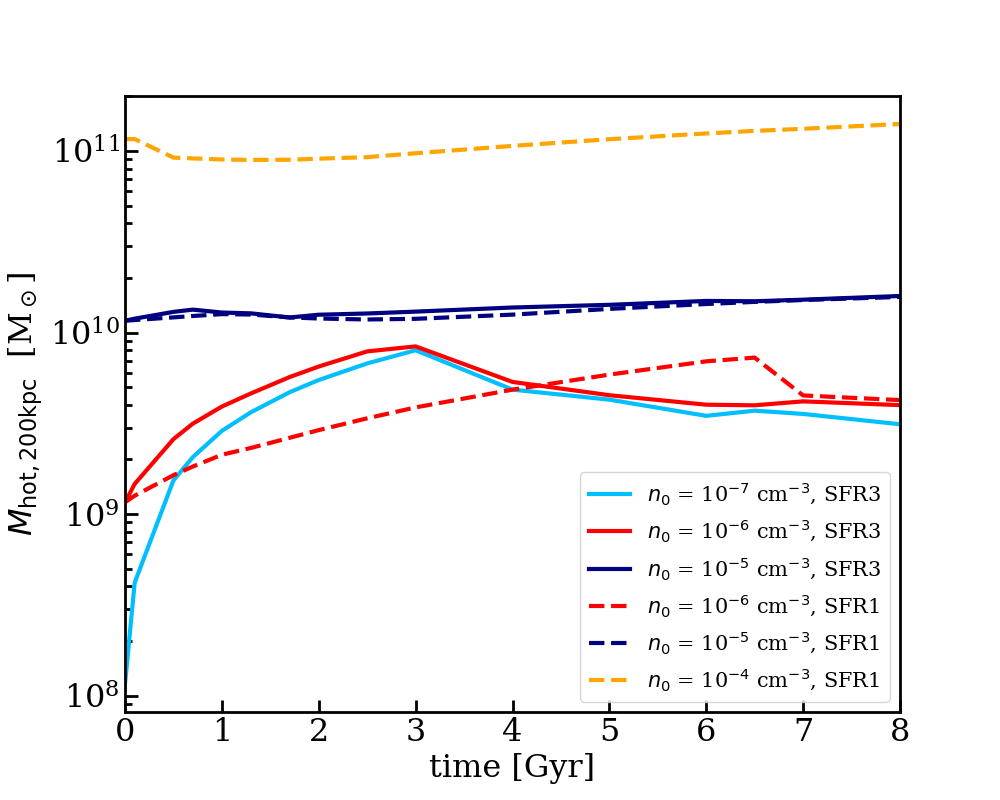}
\caption{Mass of hot gas enclosed in $R=200$ kpc as a function of time. 
}
\label{f:Mh_t_fountain_n0}
\end{center}
\end{figure}

Fig. \ref{f:Mh_t_fountain_n0} shows $M_{\rm{h}}$(R$<$ 200 kpc) for various simulations. When $n_0 \lesssim 10^{-6}$ cm$^{-3}$, the evolution is similar to what we have seen for the fiducial run.  $M_{\rm{h}}$(R$<$ 200 kpc) increases initially, until the total mass reaches 7-8$\times 10^9 M_\odot$. Then the cooling happens, taking away some mass from the hot atmosphere; after that $M_{\rm{hot,200kpc}}$ stays at around 5$\times 10^9 M_\odot$. Different SFRs only affect the time it takes to reach the maximum mass, but do not affect the value of this mass. 
For the runs with $n_0 = 10^{-5}$ cm$^{-3}$, the mass remains almost unchanged through the simulation, at 1.2$\times 10^{10}$ M$_\odot$. This is a factor of 2-3 larger than the steady state of the hot CGM with $n_0\leq$ $10^{-6}$ cm$^{-3}$. When $n_0 = 10^{-4}$ cm$^{-3}$, the outflows barely reach a few kpc, and the mass of the CGM stays close to the initial condition.

\subsection{Universal density profile of the hot atmosphere}
\label{sec:univ_den}

Because the mass of the hot atmosphere reaches a constant value, it is interesting to examine the density profile. In this section, we first use one example run to show the time evolution of the density profile. Then we show that that the density profile up to \rmax\ is independent of SFR and $n_0$. Lastly we quantify the density profile and compare to observational constraints from the MW.

\begin{figure}
\begin{center}
\includegraphics[width=0.5\textwidth]{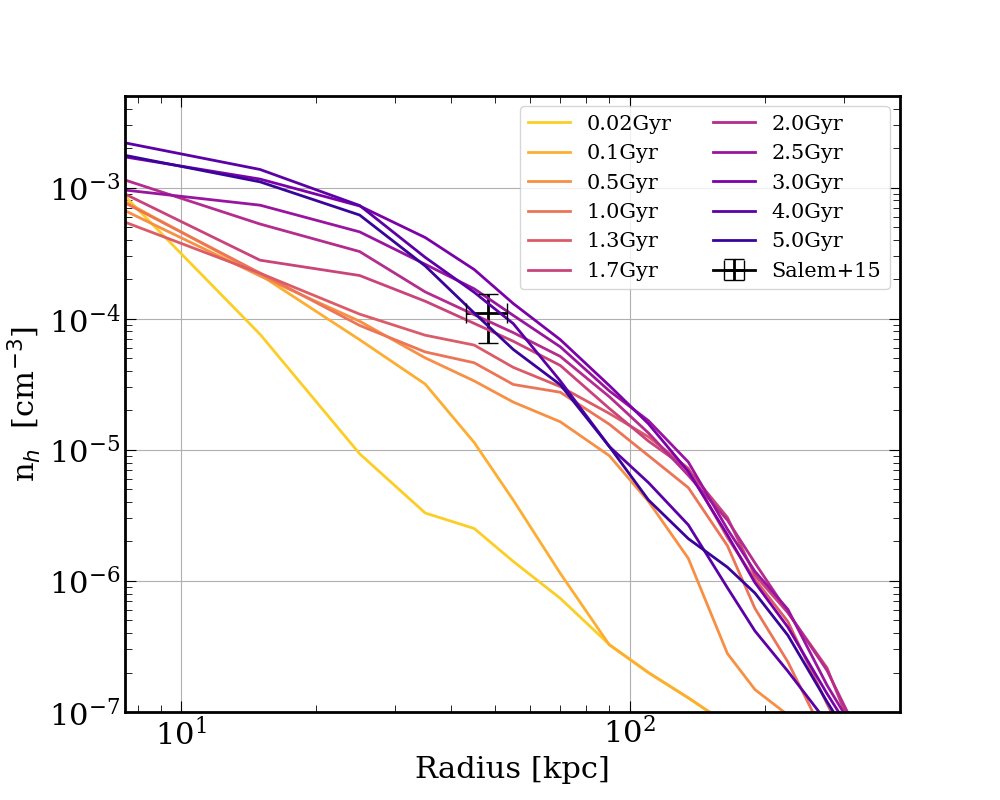}
\caption{Radial profile of hot gas density for the run 1e-7SFR3 at different times. The circumgalactic space is filled from the large radii by the SNe-driven outflows. The error bar indicates $n_h\sim 10^{-4}$ cm$^{-3}$ at $R\sim$ 50 kpc constrained by \cite{salem15} for the MW.
}
\label{f:n_r_t_1e-7_SFR3}
\end{center}
\end{figure}

Fig. \ref{f:n_r_t_1e-7_SFR3} shows the time evolution of the density profile for the run 1e-7SFR3. Since this run has the lowest $n_0$, it is easier to see the filling process of the CGM.
The density at a certain radius is the spherically-averaged value weighted by volume. Overall, the density profile does not change much after $t\gtrsim$ 1.5 Gyr. The subtle evolution after that (within a factor of 2) is similar to the evolution of the enclosed mass (Fig. \ref{f:Mh_t_fountain_n0}), that is, first an increase, followed by a slight decrease. 
Large radii ($R\gtrsim$ 100 kpc) reach their maximum density first. The error bar indicates $n_h\sim (1.1\pm0.45)\times 10^{-4}$ cm$^{-3}$ at $R\sim$ 50 kpc. This is constrained for the MW from the leading arm of the Large Magellanic Cloud \citep{salem15}. Our density profile naturally evolves to and stays around this value at $t\gtrsim$ 1.5 Gyr.

\begin{figure}
\begin{center}
\includegraphics[width=0.5\textwidth]{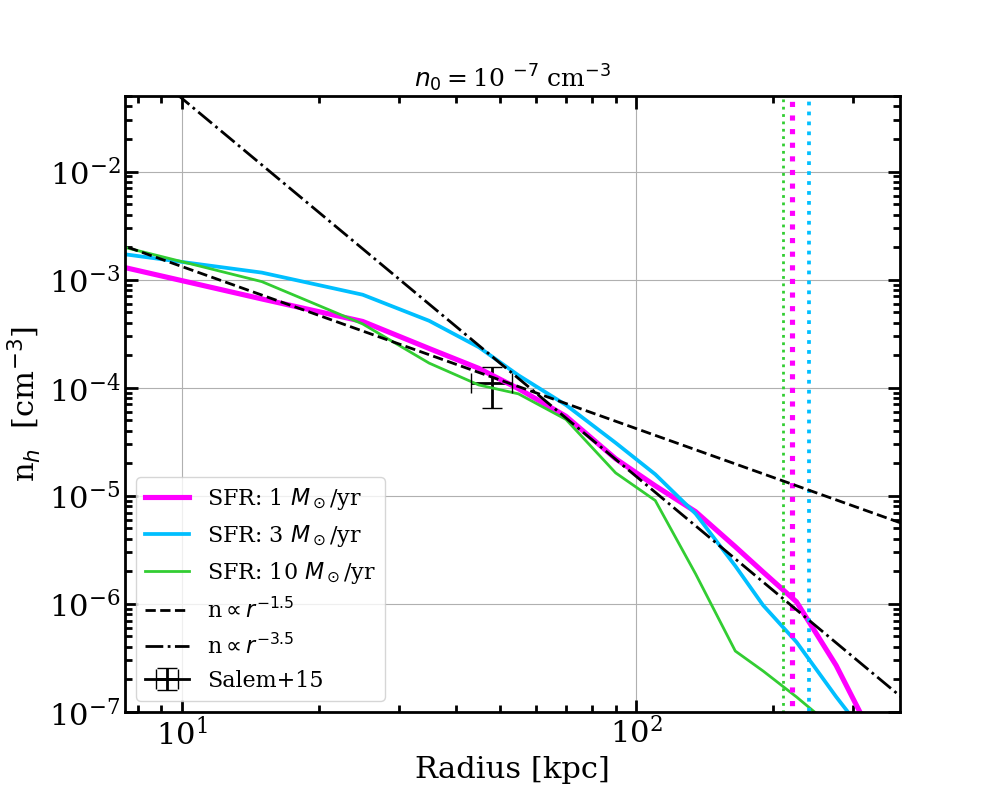}
\includegraphics[width=0.50\textwidth]{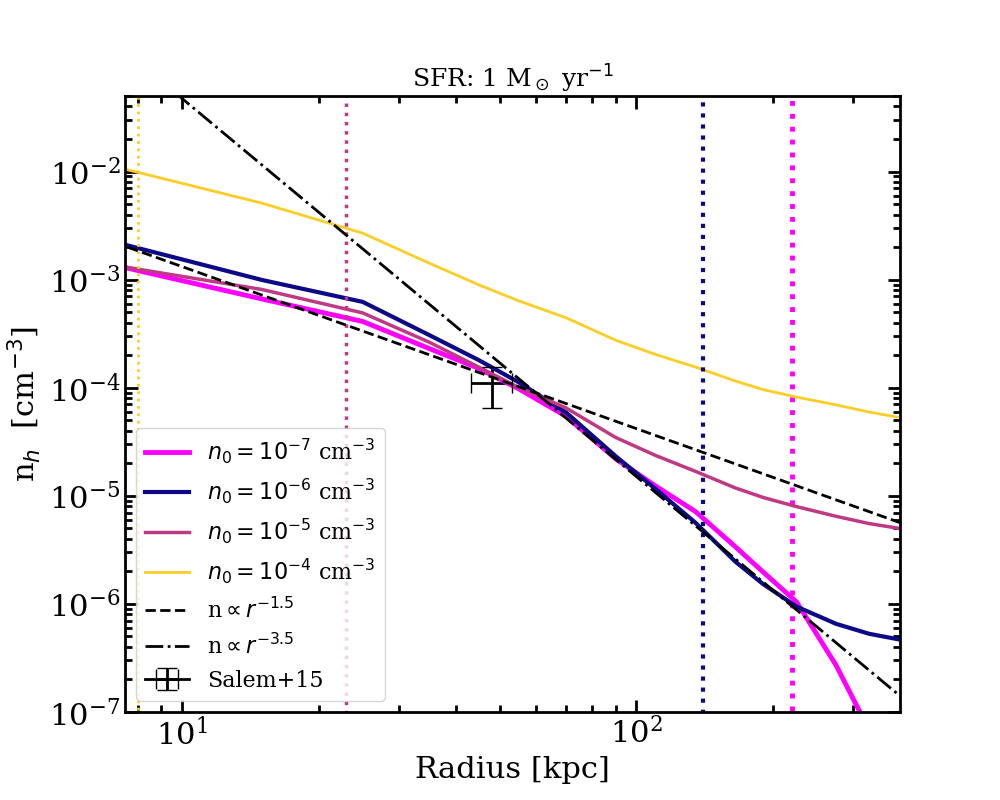}
\caption{Radial profile of hot gas density for runs with different SFRs and $n_0$. All the snapshots are taken when (6-7)$\times 10^{9}$ M$_\odot$ of hot outflows have been injected into the CGM, which is after the saturation of the hot CGM. 
Top panel: runs with SFR$=$1, 3, 10 $M_\odot$ yr$^{-1}$, respectively but with the same $n_0= 10^{-7}$ cm$^{-3}$. The vertical dotted lines of the same color indicate the final \rmax\ for each run. The density profile does not depend on SFR. 
Bottom panel: runs with different $n_0$, but with the same SFR $=$1 $M_\odot$ yr$^{-1}$. Note that the magenta curve is shown on both panels for reference, which is for 1e-7SFR1. The universal density profile is independent of $n_0$ up to \rmax\ for each run (except when  $n_0$ is very large at $10^{-4}$ cm$^{-3}$ and outflows barely get out). 
The universal profile has $n_h\propto R^{-1.5}$ at $R\lesssim$ 60 kpc, and drops faster at $n_h\propto R^{-3.5}$ at $R\gtrsim$ 60 kpc. The slope and normalization are very close to the well-constrained MW values at $R\lesssim$ 50 kpc.
}
\label{f:n_r_t_diff_SFR}
\end{center}
\end{figure}

Fig. \ref{f:n_r_t_diff_SFR} shows the radial profiles of density of the hot CGM for various runs. The top panel shows the runs with three different SFRs: 1, 3, 10 $M_\odot$ yr$^{-1}$, but with the same $n_0= 10^{-7}$ cm$^{-3}$. The bottom panel shows the runs with different $n_0$, but with the same SFR of 1 $M_\odot$ yr$^{-1}$. The snapshot of each run is taken when (6-7)$\times 10^9\ M_\odot$ outflows have been injected into the CGM. We use this criteria for comparison because these outputs are slightly after the saturation state is reached (Fig. \ref{f:Mh_t_fountain_n0}), after which point there is little change in the density profile in any run. The vertical dotted lines indicate the final \rmax\ for each run of the same color. For reference, the run 1e-7SFR1 (magenta line) is shown on both panels.

In the top panel of Fig. \ref{f:n_r_t_diff_SFR}, the resultant density profiles of the three runs look very similar. Since the only difference among the three runs is how fast the mass is being injected into the CGM, the similarity of the results indicates that the density profile does not depend on the mass injection rate (for a given set of loading factors). As long as the same amount of mass of outflows is added, the density profiles are very similar. In the bottom panel, the density profiles are the same up to their respective \rmax. The only exception is when $n_0$ is extremely large, $10^{-4}$ cm$^{-3}$, in which case outflows reach only a few kpc, and the density profile stays at the initial condition. This large density, as shown in a later section, is excluded by X-ray observations.

We plot the power-law profiles $n\propto R^{-\beta}$ to compare with the universal density profile (i.e., after the saturation is reached). Our density profile can be well approximated by a broken power law with a ``knee" at \rout $\sim$ 60 kpc, i.e.,
\begin{equation}
n_h \sim 9\times 10^{-5} \rm{cm}^{-3} \left(\frac{R}{R_{\rm{out}} }\right)^{-\beta},
\end{equation}
where
\begin{equation}
\beta \sim
    \begin{cases}
     1.5 ,\ \ \ R\leqslant R_{\rm{out}}, \\
     3.5 ,\ \ \ R_{\rm{out}}< R \leqslant R_{\rm{max}}.
    \end{cases}
\end{equation}
The density profile drops steeply beyond \rout. We show in the Appendix \ref{sec:res} the resolution test for $n_h$ at a few radii, which has good convergence.

The density profile of hot gas for the MW CGM is well-constrained at $R\lesssim$ 50-70 kpc \citep[recent summaries by][]{miller13,faerman17,bregman18}. The density at $R\lesssim$ 50-70 kpc has a power law with an index of -1.5, as constrained from the emission measure of O VII \citep[e.g.][]{henley13,miller15}, dispersion measure of pulsars in the Magellanic Clouds \citep{anderson10}, and absorption lines of O VII and O VIII \citep{gupta12,fang15,bregman18}. 
The density at a radius of 50 kpc, as mentioned before, is inferred to be (1.1$\pm$0.45) $\times 10^{-4}$ cm$^{-3}$ \citep{salem15}.
Our density profile is in excellent agreement with these constraints. At $R\gtrsim$ 50-70 kpc, the constraints are mainly from the inferred pressure of cool phases, which are likely model-dependent; different models sometimes give conflicting results \citep[e.g.][]{stanimirovic02,fox05,werk14}. Theoretical models also differ greatly at these large radii \citep{maller04,faerman17,qu18,fielding17,voit19, stern19,kauffmann19,huscher20,davies20,faerman20}. Our model, having a steep slope at $R>$\rout, gives a lower density compared to other models. Future observations of MW-mass halos, such as X-ray emission/absorption from large impact parameters and measurement of the Sunyaev-Zel'dovich effect, are necessary to robustly constrain the outer part of the diffuse CGM.

\section{Observational Signatures}

In this section, we discuss the observational signatures of the CGM, focusing on the warm-hot/hot ($T>3\times 10^4$ K) component. We will show that the observables are closely related to the underlying physical properties of the CGM.

\subsection{X-ray luminosity}

X-ray emission is observed from the ``coronae'' around disk galaxies, and the luminosity increases with the SF activity in galaxies \citep[e.g.][]{mineo12,li13}. In this subsection, we discuss the X-ray luminosity $L_X$ of the CGM in our simulations, and find that it traces the total amount of diffuse gas in the halo.

\begin{figure*}
\begin{center}
\includegraphics[width=1.0\textwidth]{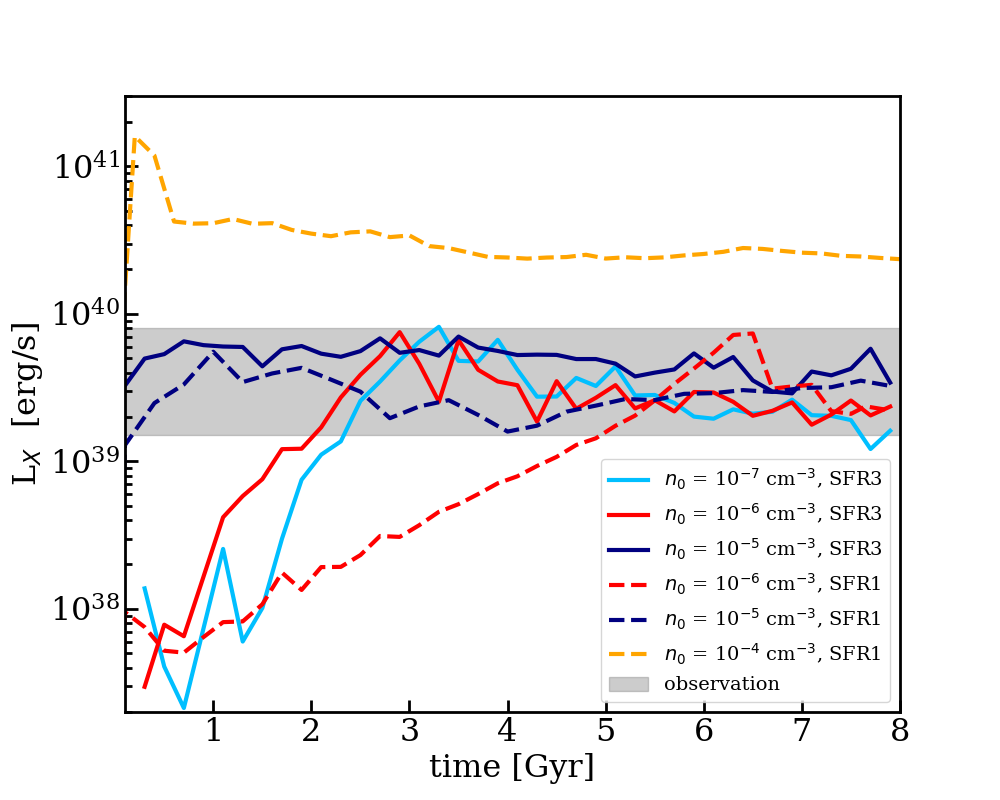}
\caption{$L_X$ as a function of time for various runs. The runs included in this plots are the same as those in Fig. \ref{f:Mh_t_fountain_n0}. The observed $L_X$ for disk galaxies with a mass and SFR of the MW \citep[e.g.][and references therein]{wang16,mineo12} is shown by the grey shaded region. When $n_0 \lesssim 10^{-5}$ cm$^{-3}$, $L_X$ settles down to the observed range. Large $n_0=10^{-4}$ cm$^{-3}$
violates the $L_X$ constraints. Also notice that the evolution of $L_X$ resembles that of $M_h$ in Fig. \ref{f:Mh_t_fountain_n0}.   }
\label{f:Lx_t_fountain_n0}
\end{center}
\end{figure*}

Fig. \ref{f:Lx_t_fountain_n0} shows the evolution of $L_X$ for runs with different $n_0$ and SFR. $L_X$ includes the X-ray emission in the energy range of 0.5-2 keV, for all gas located at 10 $\leqslant$ $R \leqslant$ 30 kpc. Most emission comes from $R\lesssim$ 30 kpc where the density is highest; $R<$ 10 kpc is excluded due to its proximity to the injection regions and large temporal variations (see Appendix \ref{sec:Lx_R} for details). $L_X$ is calculated using the yt module with the APEC emissivity model. 

The evolution of $L_X$ resembles that of $M_h$ in Fig. \ref{f:Mh_t_fountain_n0}. When $n_0$ is small, $\lesssim 10^{-6}$ cm$^{-3}$, $L_X$ first increases, followed by a peak at $t\approx$ 3-4 Gyr, and then becomes a constant at several 10$^{39}$ erg s$^{-1}$ thereafter. When $n_0 \gtrsim 10^{-5}$ cm$^{-3}$, $L_X$ stays at nearly the initial value throughout the simulation. 
The observed $L_X$ for MW-size galaxies with a SFR of 1-3 $M_\odot$/yr is 1.5-8 $\times 10^{39}$ erg s$^{-1}$
\citep[][and references therein]{wang16,mineo12}, which is shown by the grey shaded region on the plot. $L_X$ settles to the observed value if $n_0 \lesssim 10^{-5}$ cm$^{-3}$; when $n_0 = 10^{-4}$ cm$^{-3}$, $L_X$ is too large to be consistent with the observations. The observed $L_X$ corresponds to $M_h (R<200 \rm{kpc}) \sim (0.5-1.2) \times 10^{10}$ M$_\odot$. We also show in the Appendix \ref{sec:res} the resolution check for $L_X$, which shows little dependence on the numerical resolution.

In summary, $L_X$ is highly indicative of the evolution of hot diffuse CGM in the halo. The rising curve of $L_X$ corresponds to when the hot atmosphere is being filled. The peak of $L_X$ is when the hot atmosphere has largest mass and becomes saturated. After that, the hot envelope reaches a steady state, where the density profile of the hot atmosphere settles around the critical point of saturation, and $L_X$ remains unchanged. The saturated hot CGM has a $L_X$ that matches the observed value of $L_X$, if $n_0\lesssim 10^{-5}$ cm$^{-3}$. 

The X-ray luminosity only accounts for a few percent of the energy injection rate carried by the hot outflows. This poses the ``missing feedback" problem \citep{wang10}. We find that, before saturation is reached, the outflow energy is used to push outflows and pre-existing gas to large radii, i.e., the energy released from the galaxy converts into potential energy of the CGM. After saturation, most of the outflow energy is radiated away at $T< 2\times 10^6$ K. The X-rays emission at 0.5-2 keV band is mainly from gas $T\gtrsim 3\times 10^6$ K, which is above the temperature of most of the CGM. The temperature distribution of the CGM will be discussed in the next section.

\subsection{Absorption lines}

Observational studies of the CGM have relied heavily on absorption lines against background quasars \citep[e.g.][]{chen10,tumlinson11,steidel10}. This is especially useful at large radii, where the emission is too weak to detect due to the low density.

In order to gain physical intuition, we will first show the radial profiles of specific entropy, temperature and Mach number, and their differences before and after CGM saturation. Then we use the oxygen as an example and show the column density of oxygen at different ionization states, namely O VI, O VII, and O VIII, as a function of impact parameter. We emphasize how the column density distributions of these ions relate to the underlying temperature profile. We make predictions for a few other highly-ionized ions, Ne VIII, Ne IX and Mg X.
 
\subsubsection{Radial profiles of the CGM} 
 
\begin{figure*}
\begin{center}
\includegraphics[width=0.45\textwidth]{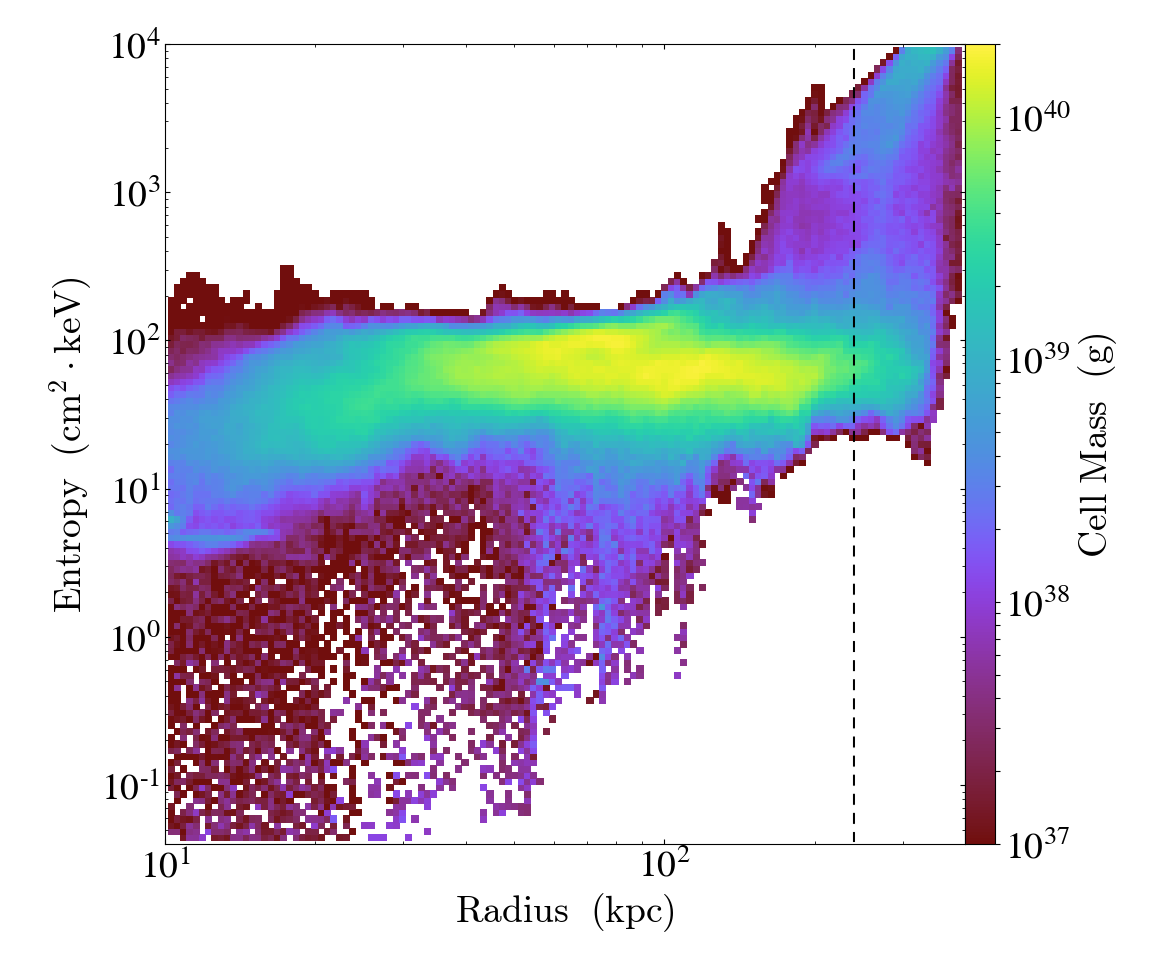}
\includegraphics[width=0.45\textwidth]{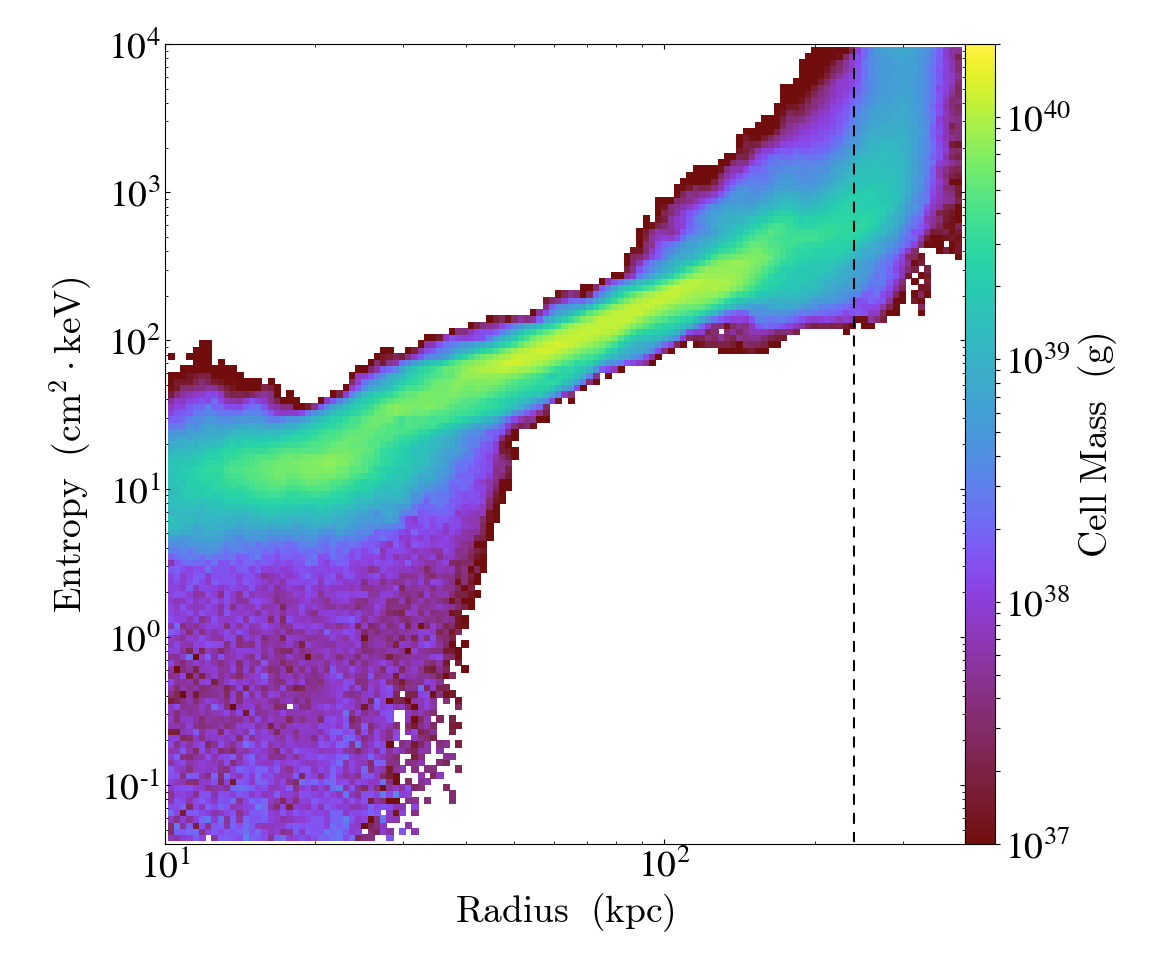}
\includegraphics[width=0.45\textwidth]{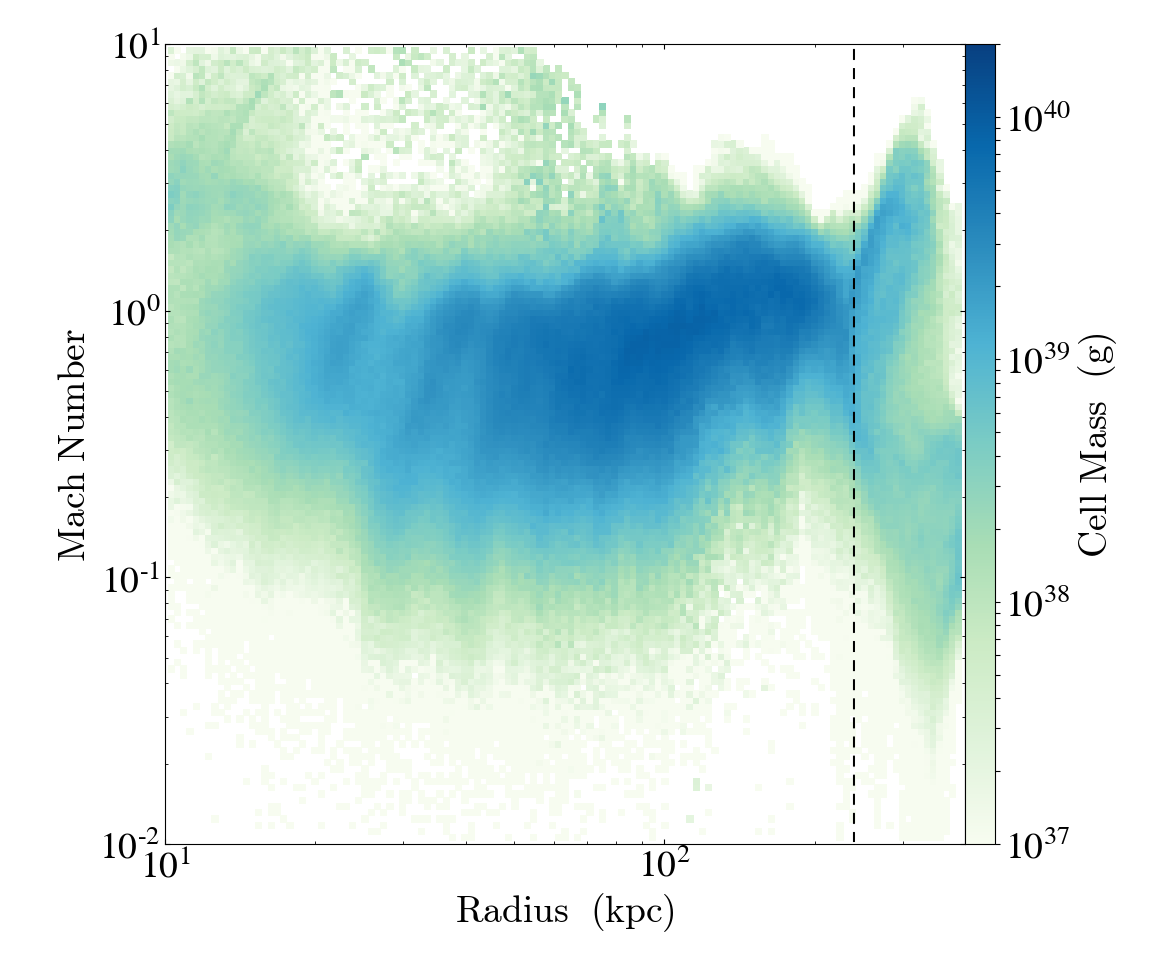}
\includegraphics[width=0.45\textwidth]{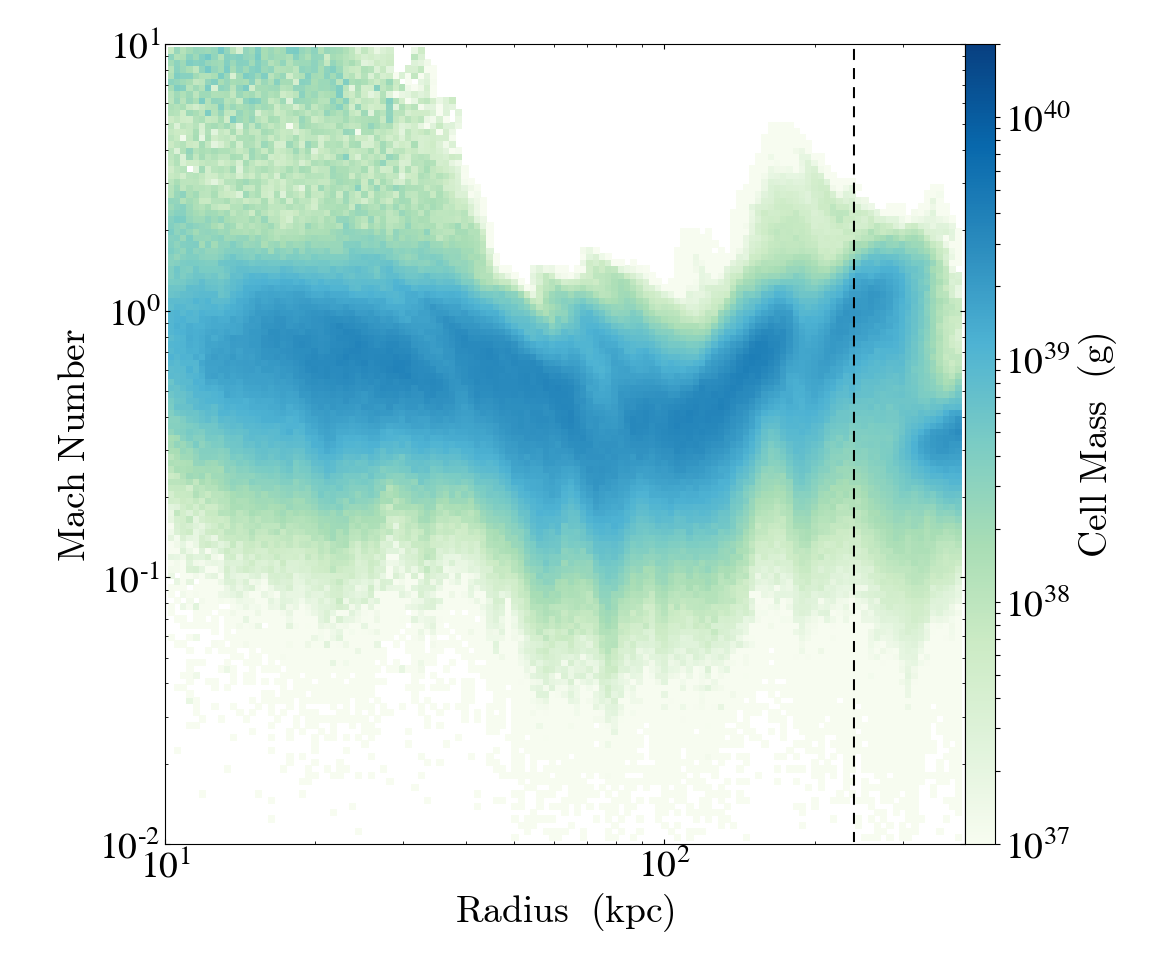}
\includegraphics[width=0.45\textwidth]{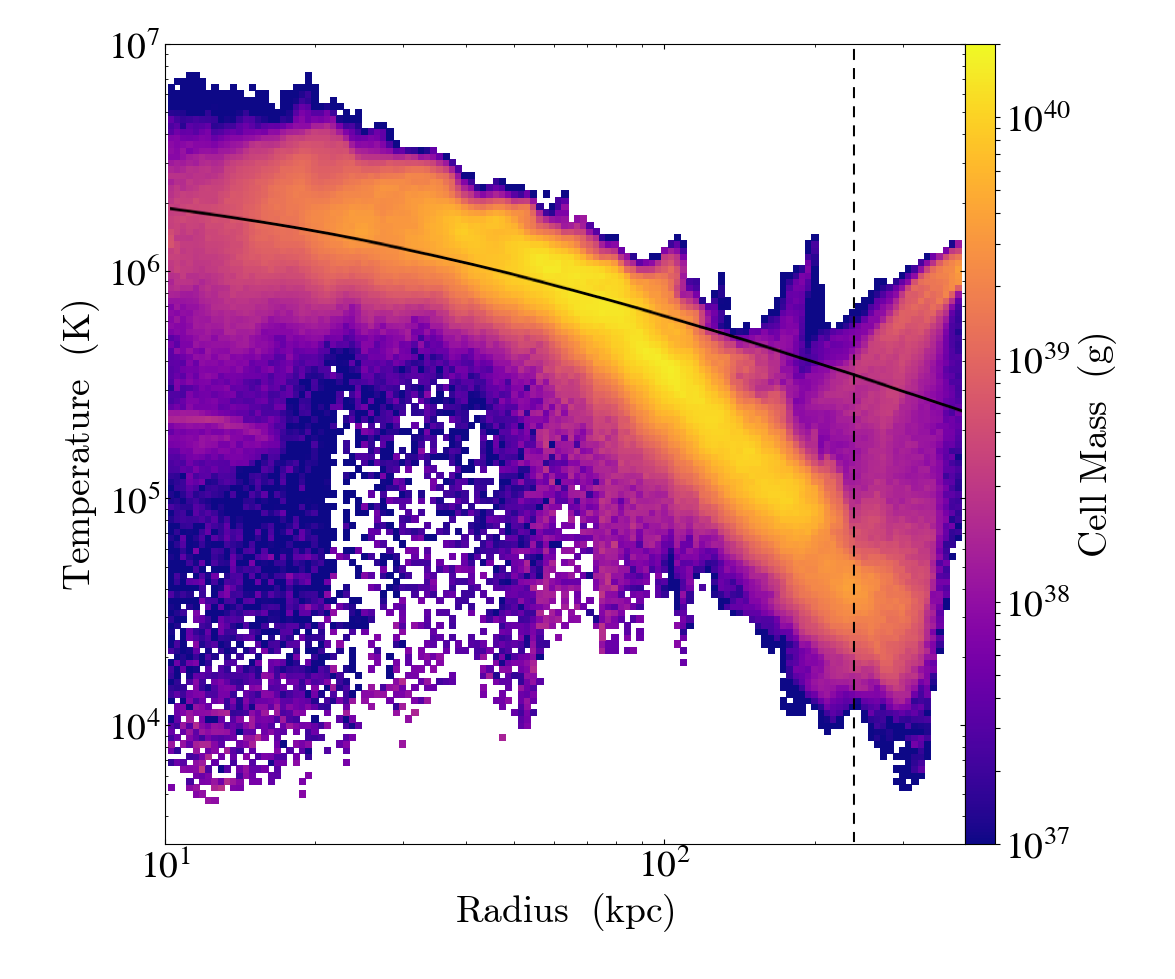}
\includegraphics[width=0.45\textwidth]{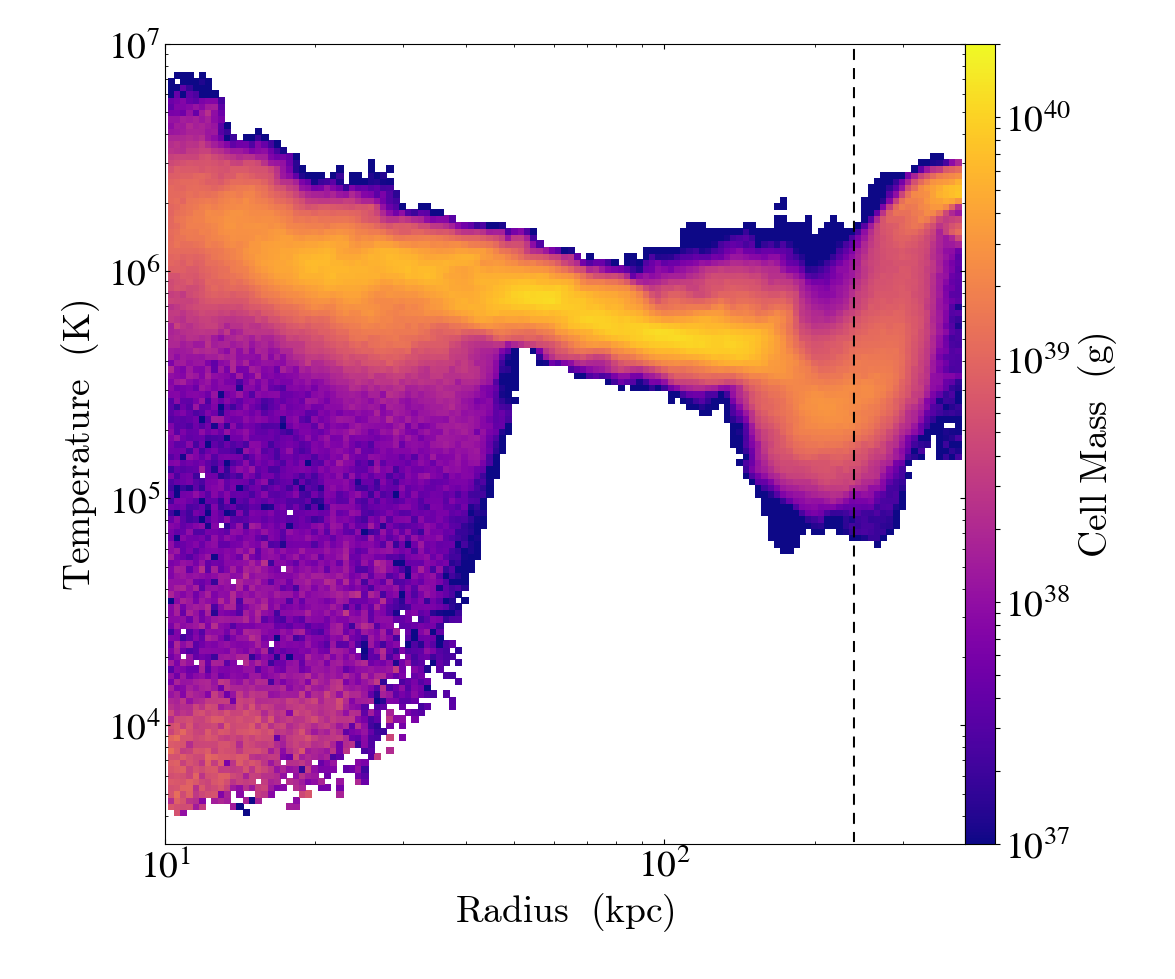}
\caption{Radial distribution of specific entropy, Mach number, and temperature for the run 1e-7SFR3. The black dashed line in each plot shows \rmax.
The left column is at 2 Gyr, prior to the saturation, and the right column is at 5.5 Gyr, after the saturation. For the left column, the specific entropy and Mach number are radially flat, and the temperature decreases with increasing radii. On the left temperature panel we plot the the gravitational potential $-\phi(r)$ with an arbitrary normalization (black line). The temperature follows $-\phi(r)$ up to \rout$\sim$ 60 kpc. For the right column, the specific entropy increases with radius, and the temperature profile is less steep. These are because of the reverse shocks induced by the collision of cool clumps at the center. The Mach number remains flat.  }
\label{f:entropy_T_r_t_1e-7_SFR3}
\end{center}
\end{figure*}

Fig. \ref{f:entropy_T_r_t_1e-7_SFR3} shows the radial profiles of specific entropy, local Mach number, and temperature for the run 1e-7SFR3. We use the lowest density run where \rmax\ is largest. Simulations with larger $n_0$ have the same profiles up to their \rmax. The snapshot on the left column is at 2 Gyr, prior to the saturated state, and the right column is at 5.5 Gyr, after saturation is reached. The black dashed line in each plot indicates \rmax. Colors indicate the gas mass.

Before the saturation state, the radial profile of specific entropy is mostly flat up to \rmax, at 80 $\pm$ 20 cm$^{2}$ keV. In other words, the fountain flows have maintained an isentropic atmosphere. Beyond \rmax, the entropy is that of the pre-existing gas, which has a value higher than the fountain flow. At $R\lesssim$ 100 kpc, there is some low-entropy gas, which is also seen in the temperature profile as gas at about $10^4$ K. This is the gas that has already started to cool, and is the ``precursor'' of the upcoming cooling event at $t=$3-4 Gyr. The flat entropy profile means that buoyancy is not at work to suppress the thermal instability in this stratified atmosphere \citep{field65,balbus89,binney09,voit17}.

The middle panel shows the profile of local Mach number, defined as the velocity of each cell divided by its sound speed. The majority of the gas between 10-200 kpc has a Mach number at 0.5-1.2, with a mean value of 0.8. This indicates that there are considerable motions in the hot CGM. The motions can also serve as ``pressure'' support of the CGM against gravity. 

The bottom panel shows the the temperature profile, which decreases with increasing radius. It has a broad range of $7\times10^5 - 5\times 10^6$ K in the inner 20 kpc, and drops to $\lesssim 10^5$ K at $R>$ 180 kpc. To understand the decreasing profile of the temperature quantitatively, we first model it as an isentropic gas in quasi-hydrostatic equilibrium with the gravitational potential. There is a simple correlation between temperature and the potential in this case. The main equations are the following:\\
(i) a quasi-hydrostatic equilibrium
\begin{equation}
    \frac{1}{\rho} \frac{dP_{\rm{tot}}}{dr} = - \frac{d\phi}{dr},
\end{equation}
(ii) A constant Mach number $\mathcal{M}$, so that
\begin{equation}
    P_{\rm{tot}} = P_{\rm{th}}  + P_{\rm{dyn}} = P_{\rm{th}} (1+ \mathcal{M}^2), 
\end{equation}
(iii) A constant specific entropy $K$,
\begin{equation}
    P_{\rm{th}} \rho^{-\gamma} = K,
\end{equation}
where $\gamma$ is the adiabatic index. Using the above three equations, we have
\begin{equation}
    \frac{\gamma (1+\mathcal{M}^2)}{\gamma-1} \frac{k_B}{\mu} T(r) = - \phi(r) + C, 
    \label{eq:T-phi}
\end{equation}
where $C$ is a constant. This means that the radial profile of the temperature follows that of the potential.

On the bottom panel of Fig. \ref{f:entropy_T_r_t_1e-7_SFR3} we plot the potential $-\phi(r)$ using a black line, with an arbitrary normalization. The temperature follows $-\phi(r)$ until \rout$\sim$ 60 kpc. Beyond that, the temperature drops faster than the potential up to \rmax. 

For an isentropic profile, 
\begin{equation}
    n \propto T ^{1/(\gamma-1)}\propto \phi^{1/(\gamma-1)}.
\end{equation}
At $R >$ the core radius of the DM (10 kpc in our case), $\phi\propto R^{-1}$, thus $n\propto R^{-3/2}$ for $\gamma=5/3$. This is consistent with the universal density profile $n\propto r^{-3/2}$ up to \rout (Fig. \ref{f:n_r_t_diff_SFR}).

The fact that beyond 60 kpc, $T(r)$ deviates from Eq. \ref{eq:T-phi} indicates the above condition (i) quasi-hydrostatic equilibrium is not satisfied (since the other two conditions are true from simulations). Both density and temperature drop quickly with radius at [\rout, \rmax]. This radial range seems to be the transitional region between the bulk of the fountains to the low-density IGM space. Interestingly, some  cosmological simulations also see decreasing temperature profile in the halo to various degrees \citep[e.g.][]{hummels13,kauffmann19,huscher20}, though the exact reason is not clear.

After the CGM saturates, the profiles of entropy and temperature have interesting shifts compared to the pre-saturation state, as seen in right column of Fig. \ref{f:entropy_T_r_t_1e-7_SFR3}. 
The inner 30 kpc of the entropy profile is flat for the hot phase, at 10-20 cm$^2$ keV. This is the region where the cool phase forms, evidenced by the low-entropy patches on the lower-left part of the diagram. This is consistent with the conjecture of \cite{maller04}, that the cooling instability leaves a constant entropy for the remaining hot atmosphere, and is also seen in cosmological simulations \citep[e.g.][]{huscher20}. The flat entropy profile allows future thermal instability to proceed.
The temperature profile becomes less steep at large radii. The rising profile of entropy and the less steep profile of temperature are due to outward shocks. These are reverse shocks induced by collisions of cool gas after they fall to the center of the halo with velocities of several hundred km s$^{-1}$, as a result of the colliding cool clumps at the center of the halo. In fact, some shock fronts are visible on the temperature profile, e.g. a temperature jump at about 160 kpc. These shocks increase the entropy and temperature at larger radii. Our simulations do not have an ISM to start with, but if one was included, the reverse shock should still be present, since cool CGM gas falling at several hundred km s$^{-1}$ would collide with the ISM in the galaxy and induce strong reverse shocks into the CGM.

The radial profile of the Mach number, on the other hand, only shows minor changes over time. The majority of the gas has $\mathcal{M}=$0.6$\pm$0.3 after the condensation, in comparison to 0.9$\pm$0.3 before the condensation. The motions are maintained by the constant injection of the energy from outflows. The high-Mach number ($\gtrsim$ a few) patches at $R\lesssim$ 30 kpc are from the cool clumps falling toward the galaxy ballistically. 

\begin{figure}
\begin{center}
\includegraphics[width=0.48\textwidth]{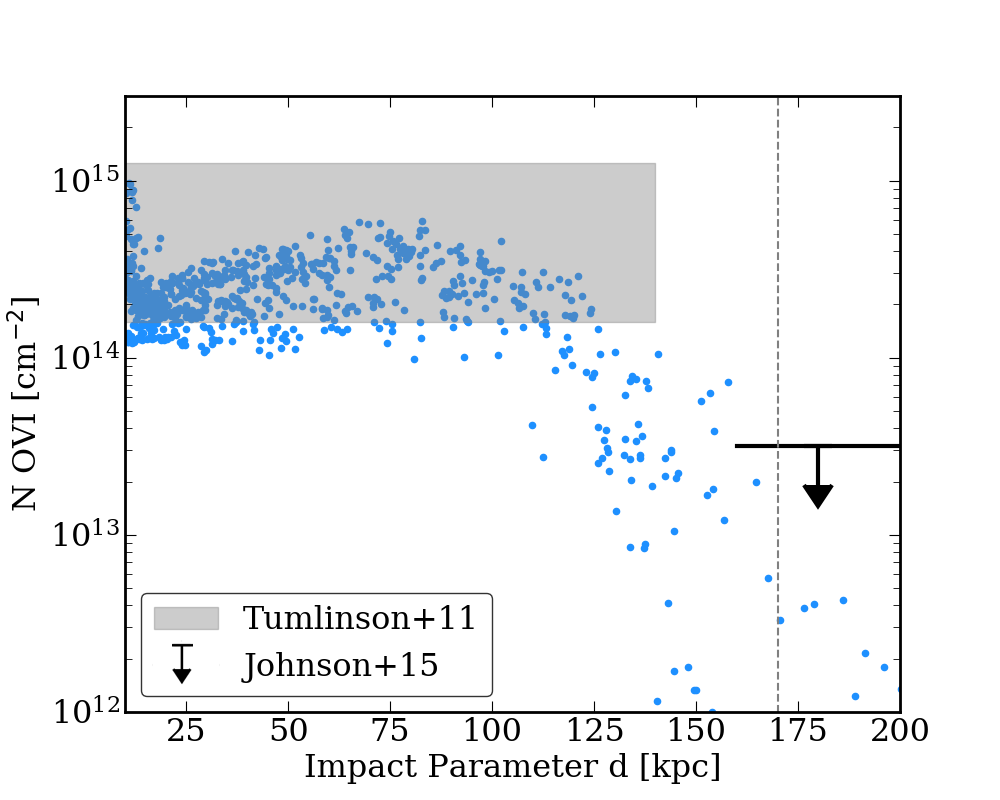}
\includegraphics[width=0.48\textwidth]{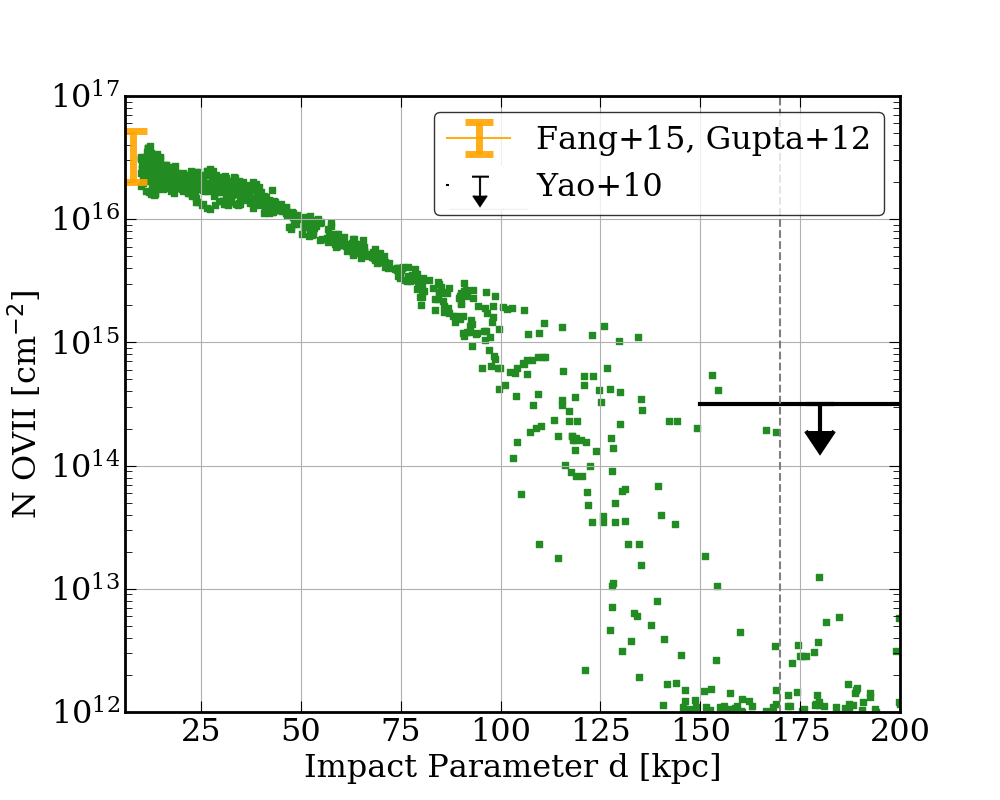}
\includegraphics[width=0.48\textwidth]{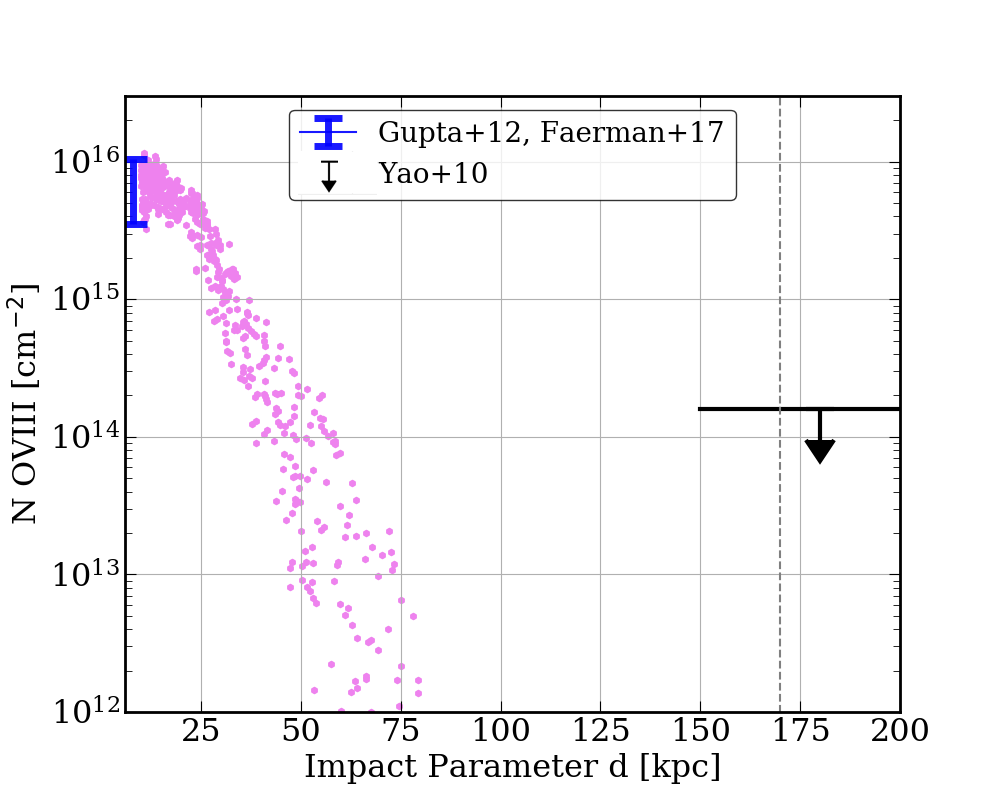}
\caption{
Column densities $N$ (O VI) (top panel), $N$(O VII) (middle), and $N$ (O VIII) (bottom) as a function of impact parameter $d$. This is at 2.5 Gyr for the run 1e-7SFR3. The points are random sight lines from the simulation, assuming collisional ionization equilibrium. The vertical dashed line indicates the radius beyond which $n<10^{-5.5}$ cm$^{-3}$, where non-CIE effects can be important.
The shaded region, error bars and upper limits are observational constraints. The time evolution of  N (O VI) is shown in Fig. \ref{f:N_OVI_t}, whereas $N$ (O VII) and $N$ (O VIII) change very little with time.  
}
\label{f:N_O678}
\end{center}
\end{figure}

While Fig. \ref{f:entropy_T_r_t_1e-7_SFR3} shows two representative snapshots, we note that there is little time evolution of the profiles in the time windows of 1-3.5 Gyr (after the outflows reach \rmax\ and before the saturation) and $\gtrsim$4.5 Gyr (after the reverse shocks propagate through the CGM), respectively. The temperature of the medium determines the ionization state of ions when the ionization is caused by collision. The temperature decreases radially, meaning that ions are at progressively lower ionization states at larger radii. We now discuss column densities of O VI, O VII, and O VIII, and connect them to the underlying physical properties of the CGM, i.e., the radial profiles discussed above.

\vspace{0.1in}

\subsubsection{O VI}

When calculating the column densities, we assume a number density ratio O/H$=4.9\times 10^{-4}$ for the solar metallicity \citep{asplund09}. Note that there is a factor of 2 for the uncertainty of the ion abundances. An additional but smaller uncertainty comes from our model input of the ISM metallicity, which is 0.8 $Z_\odot$, though this value evolves across cosmic time. In this paper we also assume collisional ionization equilibrium (CIE), and use the ionization table from \cite{mazzotta98}. 
We recognize that non-equilibrium effects and photoionization can leave the ionization states different from what CIE dictates, particularly in low-density regions \citep[e.g.][]{breitschwerdt94,oppenheimer13,corlies16,stern18}. But since the cosmic ionizing background is redshift-dependent \citep[e.g.][]{haardt12}, has great uncertainties \citep{kollmeier14,shull15}, and can also be stochastic \citep[e.g.][]{oppenheimer18b}, we postpone the inclusion of these effects to future work. While presenting results under CIE, we will estimate the possible effects when relevant, which is generally small except for at large radii. 
In the plots, we mark the radius at which the density is below $10^{-5.5}$ cm$^{-3}$, where these effects can be important.

\begin{figure}
\begin{center}
\includegraphics[width=0.5\textwidth]{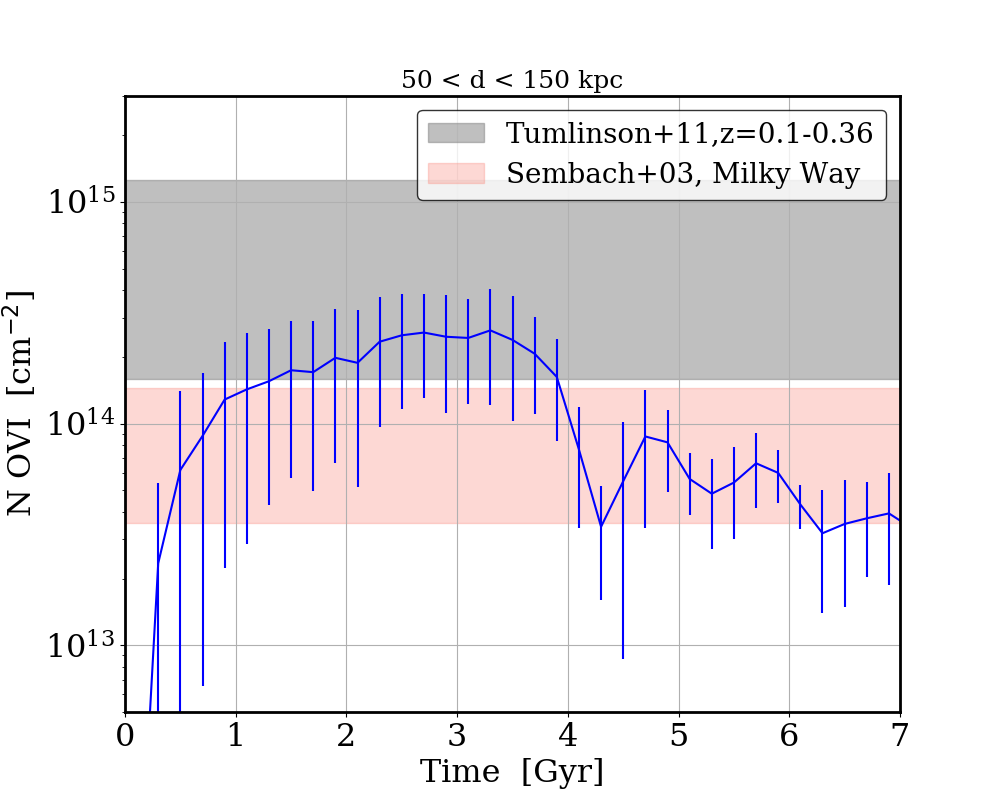}
\caption{Evolution of N (O VI) at impact parameter $50 <d< 150$ kpc for the run 1e-7SFR3. The solid line and the error bars indicate the mean and standard deviation of N (O VI) from different sight lines. The grey shade indicates the COS-Halos observations at redshifts 0.1-0.36 (a few Gyr in look-back time) \citep{tumlinson11}, whereas the pink shade indicates the observed N (O VI) from the MW halo \citep{sembach03}. Our simulation reproduces this evolution.}
\label{f:N_OVI_t}
\end{center}
\end{figure}

Fig. \ref{f:N_O678} shows the column densities $N$ of O VI, O VII and O VIII as a function of impact parameter $d$ from the the galaxy, viewed from outside the halo. The points are from randomly selected sight lines. Due to the spherical nature of the fountain, the viewing angle does not make a difference. The snapshot is taken at 2.5 Gyr for the run 1e-7SFR3. We choose 2.5 Gyr because it is shortly before saturation of the halo (which occurs at $\sim$3.5 Gyr). However, the difference between the unsaturated state and saturated one is only significant for O VI, which we will discuss in Figure \ref{f:N_OVI_t}.

N (O VI) is flat at $d\lesssim 120$ kpc and drops sharply beyond that. The COS-Halos survey has observed O VI around MW-mass galaxies (log $M_*/M_\odot \sim$ 9.5-11) at redshifts $=$0.1-0.36, out to $d \sim$ 150 kpc \citep{tumlinson11}. The detection rate around star-forming galaxies is very high, in contrast to quiescent galaxies. The observed range of $N$ (O VI) $\approx 10^{14.2-15.1}$ cm$^{-2}$ is marked by the shaded region in the top panel of Fig. \ref{f:N_O678} \citep{tumlinson11}. The eCGM survey, which covers galaxy masses of log $M_*/M_\odot \sim$ 9-11, extended to much larger $d$, and found that beyond about 160-200 kpc, N (O VI) is much smaller with only upper limits \citep{johnson15}. Our results show remarkable agreement with these observations.

We can gain insight into O VI observations by examining the physics of O VI-bearing gas in our simulations. 
O VI exists in a very narrow range of temperature $10^{5.5\pm0.1}$ K. Given our temperature profile, the O VI-bearing gas is located in a shell at $R=100 -150$ kpc. This relatively low-temperature shell forms as the hot outflows expand in the gravitational potential. 
Gas at $R=100 -150$ kpc has a low density of about $10^{-5}$ cm$^{-3}$ (Fig. \ref{f:n_r_t_diff_SFR}). This means that the O VI-bearing gas has a long cooling time of 3-5 Gyr. Therefore, the O VI-shell is a relatively long-lasting structure, in contrast to the common assumption that O VI-bearing gas is cooling rapidly \citep[e.g.][]{heckman02,faerman17}.

Indeed, we find the N (O VI) remains at the state in Figure \ref{f:N_O678} for about 3 Gyr, until after the cooling begins. Fig. \ref{f:N_OVI_t} shows the mean N (O VI) as a function of time. The error bars indicate the standard deviation of N (O VI) from different sight lines. The calculation includes sight lines at 50 $<d<$ 150 kpc; the inner 50 kpc is excluded because after cool clumps form, there is O VI cospatial with the cool gas (which exists in the inner 50 kpc). Some of the O VI comes from the interface between the cool phase and hot surroundings and is not well resolved in the simulations, thus we do not consider O VI at $R<$ 50 kpc as predictive results.  
The mean N (O VI) reaches above  10$^{14}$ cm$^{-3}$ at $t\approx$ 0.8 Gyr, when $R\gtrsim$ 100 kpc is filled to $10^{-5}$ cm$^{-3}$ (Fig. \ref{f:n_r_t_1e-7_SFR3}). It lasts until $t\approx$4 Gyr, after which the mean N (O VI) decreases by a factor of a few to  $10^{13.5-13.9} $ cm$^{-3}$. The decrease is because the reverse shock, which happens after cool clumps fall to the center, heat the outer CGM (right panel of Fig. \ref{f:entropy_T_r_t_1e-7_SFR3} ). 

The grey shade in Fig. \ref{f:N_OVI_t} again indicates the COS-Halos observations \citep{tumlinson11}, whereas the pink shade indicates the observed N (O VI) from the MW halo \citep{sembach03}. For the MW, we use the column density from the O VI category with no cool gas counterpart \citep[table 3 of][]{sembach03}, because O VI can also arise from the interface between the cool gas and the hot. The shaded regions indicate upper limits since these are detected  N (O VI) -- some sight lines have no detection. 
In addition, because the sight lines originate from within the galaxies, the column densities would be approximately half compared to the sight lines from outside the halo passing through the center. We therefore multiply the observed N (O VI) of MW by a factor of 2 for a fairer comparison with the the COS-Halos results (this, of course, assumes that the O VI distribution is spherically symmetric). In general, sight lines from the galaxy can only be compared to external sight lines which also pass through the galaxy. But since the COS-Halos reveals a flat N (O VI) as a function of the impact parameter, it makes little difference whether sight lines pass through the halo center or not. Overall, the COS-Halos observations, which are at redshifts 0.1-0.36, i.e., a few Gyr in look-back time, have an N (O VI) a factor of a few larger than the MW values. If indeed there is such a time evolution of O VI column density for the MW, our simulation interestingly reflects this evolution.

Including photoionization from the cosmic ionizing background would not change our results significantly.  Gas at $10^{5.5}$ K has a density of about $10^{-5}$ cm$^{-3}$, which means the photons can lower the fraction of O VI at this temperature \citep[e.g.][]{faerman20}; but at the same time, gas at lower temperatures will have a non-negligible fraction of O VI. Using the ratio of OVI due to photo-ionization versus collisional ionization from \citet[][their Fig. 5]{faerman20}, we estimate including photoionization would change the results by no more than a factor of 2. 
Another effect that can contribute to the O VI fraction is a dynamical non-equilibrium effect: since the $10^{5.5}$ K gas comes from the expansion of hotter outflows, the ions can be frozen at higher ionization states for a recombination timescale. The radiative recombination timescale for O VII at $10^{-5}$ cm$^{-3}$ is about a Gyr, smaller by factor of a few than the ``life-time'' of O VI in our simulation. So this non-equilibrium effect would not change the results significantly.

So far we have discussed the example case of 1e-7SFR3. We now address how the input parameters of SFR and $n_0$ affect N (O VI) and its evolution. When the SFR is lower, the pre-saturation O VI-shell lasts longer, since it takes more time for the condensation and reverse shock to happen. On the other hand, if SFR is even larger, the pre-saturation O VI-shell would last too briefly. 
If the initial density is large, $n_0 > 10^{-6}$ cm$^{-3}$, hot outflows will not reach $R> 100 $ kpc (Fig. \ref{f:rmax_diff_n0_SFR}), and thus the CGM beyond \rmax\ has a temperature of $10^6$ K as the initial condition. Therefore temperature at all radii is $>10^{5.5}$ K and N (O VI) does not exceed 10$^{14}$ cm$^{-3}$. So to have N (O VI) 10$^{14}$ cm$^{-3}$ as COS-Halos survey indicates, one needs a relatively small $n_0 \lesssim 10^{-6}$  cm$^{-3}$ and a SFR $\lesssim$ 3 \msunyr. 

We caution that O VI may occur at different types of locations. One is the large-scale volume-filling gas. This is the case for our O VI-shell. The others are generally in thin layers, such as interfaces between the hot ($\gtrsim$10$^6$ K) and cool ($\lesssim 10^4$ K) gas due to mixing and/or thermal conduction \citep[e.g.][]{gnat10,kwak10,li17b,ji19}, or cooling of hot gas that passes the intermediate temperature \citep[e.g.][]{heckman02}.
While the latter cases usually occur on small scales associated with (the formation of) cool gas, the first scenario can be a large-scale feature independent of the cool clumps. To be clear, O VI shown in this paper is from the volume-filling $10^{5.5}$ K gas. The small-scale O VI should exist in nature but cannot be robustly quantified from our macroscopic simulations, since the interface is not resolved.

From recent cosmological simulations, O VI production is sensitive to the feedback prescription used \citep[e.g.][]{hummels13,nelson18}. N (O VI) for the halos of MW-like galaxies is generally under-produced compared to the COS-Halos survey by a factor of 2-3 \citep{hummels13,suresh17,gutcke17,oppenheimer18}. These simulations also show considerable variations of the radial profile of N (O VI), ranging from a centrally-peaked configuration \citep{suresh17,nelson18} to a flatter profile \citep{ford16,gutcke17,oppenheimer18}. We note that \cite{oppenheimer18} find that in EAGLE simulations N (O VI) in the MW-like halos mainly comes from warm-hot gas at large radii, similar to our results.

\subsubsection{O VII and O VIII}

In this subsection, we focus on the higher ionization states of oxygen, O VII and O VIII. Their column densities as a function of the impact parameter are shown in the lower two panels of Figure \ref{f:N_O678}. N (O VII) declines with increasing $d$; at $d> 100$ kpc, the decline becomes much steeper. Because O VII exists for a broad temperature range of $10^{5.5}-10^{6.5}$ K, O VII-bearing gas actually fills the volume at $R\lesssim 100$ kpc. The decline at $d>$100 kpc is mainly due to the decline of density with increasing radius. In contrast, O VIII only exists at small radii below tens of kpc since it requires the largest temperature at $10^{6.2-6.7}$ K. 

O VII and the O VIII absorption lines are detected so far only for the MW (from many sight lines), and we show the observed values using the error bars on Fig. \ref{f:N_O678} \citep{fang15,gupta12}. Since we show the simulation results using sight lines from outside of the halo, same as for O VI, we apply a factor of 2 increase of the observed range to account for the location of observers from inside the MW. For O VIII, we use the column densities given in \cite{faerman17}, which were converted from the equivalent width reported in \cite{gupta12}.
For external galaxies, only upper limits are available for N (O VII) and N (O VIII) from X-ray stacking \citep{yao10} at large impact parameters, which are shown by the arrows in the figure. Our results show excellent agreement with these observational constraints. 

For the time evolution, both N (O VII) and N (O VIII) change little with time, unlike O VI. This is because they are from higher temperature gas, which is located in the inner halo, where the temperature profile does not change much over time. For the same reason, N (O VII) and N (O VIII) distributions are not sensitive to SFR, or $n_0$ (except when $n_0$ is very large $\gtrsim 10^{-4}$ cm$^{-3}$, in which cases outflows barely get into the CGM). Also, including photoionization will have a very minor effect on N (O VII) and N (O VIII) since the gas has higher densities. 

We comment briefly on the velocities of the gas that bears O VI, VII, and VIII, which are reflected in the non-thermal broadening of absorption lines. We have found a flat radial profile of Mach number of 0.8$\pm$0.3, which only shows a slight decline over several Gyr, so we will use these numbers as a simple estimate for the velocities. For O VI, a Mach number of 0.8 and a temperature of $10^{5.5}$ K gives a 3D velocity of 52 km s$^{-1}$. This is consistent with the observed  non-thermal broadening of $\approx 40-50$ km s$^{-1}$ found in COS-Halos surveys \citep{werk16}.
For O VII and O VIII, there are no observational constraints on the line widths yet. We therefore predict them to be 110$\pm$ 20 km s$^{-1}$, and 180 $\pm$30 km s$^{-1}$, respectively. With future X-ray missions such as Athena, Lynx and HUBS, which have unprecedented spectral resolution at the eV level, the motions of the hot gas can be constrained. Note that we do not have CGM rotation in our modeling, which can contribute to the velocities as well \citep{hodges-kluck16}.

Finally, we checked that the numerical resolution does not change the results for the column densities of O VI, VII, or VIII, since they are all well resolved structures (see Appendix \ref{sec:res}). This is, however, generally not true for O VI located at the interface between hot and cool phases, which is not resolved. 

\vspace{0.3in}

The above observables, $L_X$, column densities of O VI, O VII, and O VIII are the currently available observational constraints for the (warm-)hot CGM of MW-like galaxies. Before we move on to predictions for future observations, we would like to first summarize how these observables place constraints on our model inputs, i.e., $n_0$ and SFR (we do not regard the loading factors of outflows as free parameters since they are taken from small-box simulations). First we note that the initial conditions cannot simultaneously match all the four constraints. In particular, N (O VI) is significantly under-produced since all the halo gas is at $10^6$ K. It is then not surprising that N (O VI), among the four observables, puts the tightest constraints on the model inputs. Successfully reproducing N (O VI) requires $n_0\lesssim 10^{-6}$ cm$^{-3}$ and SFR $\lesssim$ 3 M$_\odot$/yr, whereas $L_X$, N (O VII), N (O VIII) only requires the initial $n_0\lesssim 10^{-5}$ cm$^{-3}$. 
The small initial density $n_0\lesssim 10^{-6}$ cm$^{-3}$ implies a tenuous halo as an initial condition ($\lesssim$ 1\% of the cosmic baryons are in the halo). How such a condition can be satisfied and its implications for the MW formation will be discussed in Section \ref{sec:discuss}.
Also note that the runs that yield a CGM matching all four constraints, the observables come from gas at $R<$\rmax, that is, within the outflow-turned-fountain region, rather than from the outer layer which remains the initial condition. This means that these observables are the model outputs rather than inputs.

\subsubsection{Ne VIII, Ne IX, and Mg X}

\begin{figure}
\begin{center}
\includegraphics[width=0.5\textwidth]{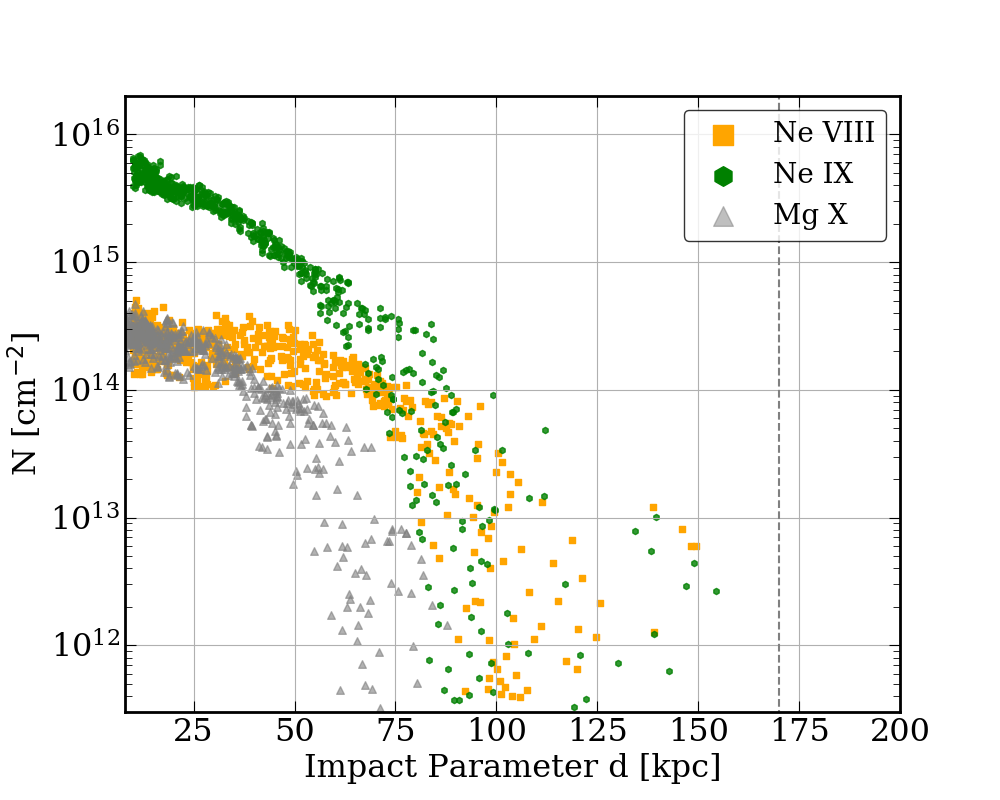}
\caption{Same as Fig. \ref{f:N_O678} but for Ne VIII, Ne IX, and Mg X. Time evolution is insignificant.}
\label{f:N_neon_mg}
\end{center}
\end{figure}

We show predictions for various ions in the warm-hot regime, Ne VIII, Ne IX, and Mg X, for which data has begun to be collected and will be plentiful with future X-ray and UV telescopes \citep{tripp11,meiring13,qu16,burchett19,das19}. A further motivation is that recent analytic models for the MW CGM by \cite{voit19} and \citep{faerman20} produce the observed $N$ (O VI), $N$ (O VII), and $N$ (O VIII) and $L_X$, \textit{with similar underpinnings of a radially-decreasing temperature profile} to our simulation results. Their separate quantitative assumptions accounting for such a temperature profile are a bit different: \cite{voit19} assume the CGM is in the precipitation limit (with additional local temperature fluctuations), whereas \cite{faerman20} assumes an isentropic CGM. 
Nevertheless, we all find that ions with higher ionization states are more concentrated in the center, and that O VI exists mainly at large radii. They both make predictions for $N$ (Ne VIII) and \cite{faerman20} also include $N$ (Mg X). So for a more robust comparison, we do the same exercise. 

We use the chemical abundances in \cite{asplund09} and the CIE table for ion fractions from \cite{mazzotta98}. Fig. \ref{f:N_neon_mg} shows column densities of these ions. Consistent with what we have found for oxygen, Ne IX is more concentrated in the center than Ne VIII; Mg X has the highest ionization temperature and is the most centrally-concentrated of the three. Also, similar to O VII and O VII, the time evolution of the column density distributions are insignificant since they all come from the inner CGM. Notably, \cite{das19} recently reported a detection of Ne IX in the MW halo, with $N$ (Ne IX) $=$ 9.3$^{+4.7}_{-3.4} \times 10^{15}$ cm$^{-3}$. At the smallest impact parameter our result is consistent with this detection.

Our predictions for $N$ (Ne VIII) and $N$ (Mg X) agree well with \cite{faerman20} at $R\lesssim$ 75 kpc. Our $N$ (Ne VIII) also agrees with the values at $d=$ 50 kpc predicted in \cite{voit19}. This corroborates the similarities of our radially-decreasing temperature profiles in the inner halo. Differences exist at large radii, which is mainly because they have a larger density than ours at large radii, and also because \cite{faerman20} includes photoionization effects. 
We will discuss in more detail about the comparisons in Section \ref{sec:discuss}.

\section{Discussion}

\label{sec:discuss}

In this Section, we discuss the implications of our results, possible effects of missing pieces, and comparisons to other work.

\subsection{Sensitivity to variations of loading factors}

In this paper, we use a specific set of loading factors of mass, energy, and metals for hot outflows. Different modeling of small-box simulations sometimes leads to different results. How would the CGM change if we use a different set of loading factors from small-box simulations? 

First, we point out that from the small-box simulations, the three loading factors of hot outflows are tightly correlated \citep[][]{li20}. Note that the simulations compiled by \cite{li20} use different numerical codes and detailed physics included are not the same, either. Nevertheless, consistent results emerge. Specifically,
\begin{equation}
    \frac{\eta_{E,h}}{\eta_{m,h}} \sim 0.8  \dot{\Sigma}_{\rm{SF}}^{0.2}, \\ \ \ \ \
    \frac{\eta_{E,h}}{\eta_{Z,h}} \sim 0.5,
\end{equation}
where \sigSFR\ is in units of \msun\ yr$^{-1}$ kpc$^{-2}$. These correlations hold for over 4 orders of magnitude in \sigSFR. 
The tight correlation indicates that for simulations that have modeled similar SF conditions but obtained different loading factors, the three loading factors differ by a constant factor. For example, \cite{kim18} modeled the solar neighbourhood condition, and have \etamh$=$0.18, \etaEh$=0.044$, \etaZh$=$0.078 (\etaZh\ is from Kim et al., in prep, private communication).  These loading factors are lower than what we use in this paper by a factor of 5.6, 6.8, 6.4, respectively\footnote{For those curious about a factor of 6 difference, this is mainly because of the different SN scale heights adopted, which have little observational constraints. A larger scale height leads to more SNe exploding above the ISM layer and thus larger loading factors of hot outflows. See \cite{li17a} and \cite{kim18} for more detailed discussions.}. 
Since the outflow fluxes scale with SFR times the loading factors, adopting their loading factors is essentially similar to using our loading factors but with a SFR 6 times lower. From Section \ref{sec:univ_den}, we see that this would make the filling of the CGM slower by a factor of 6, but does not make a qualitative difference on the CGM properties. 

Cosmic rays can also help launch galactic outflows. Recent simulations show that cosmic rays can significantly change the phase structure of the ISM and outflows, driving outflows that are predominately in cool phase and are at lower speed than the SNe-driven hot outflows \citep{simpson16,girichidis18}. That said, the studies of cosmic rays are still at early stages, and many questions and issues need to be addressed. For example, the ability of cosmic rays to drive outflows depend sensitively on the detailed physics/parameters, such as transport mechanism, diffusion coefficient, etc. \citep{salem14,wiener16,farber18}, which have large uncertainties. Notably, including cosmic rays generally leads to an ISM and outflows that are short of hot gas, which is unsupported by X-ray observations \citep{peters15}. Hence, while we acknowledge that cosmic rays may play a role in affecting loading factors of outflows, their true impacts remain to be understood and quantified.

\subsection{Missing baryon problem}

Most of the cosmic baryons are not in galaxies \citep[e.g.][]{fukugita98,bell03,guo10,moster10}. 
For the MW, if assuming the baryon to DM ratio is the cosmic value, the baryonic mass within a $10^{12}$ $M_\odot$ halo is around $1.5\times 10^{11}$ $M_\odot$. 

Roughly half of this baryonic mass does not reside in the MW or its satellite galaxies. These missing baryons are likely to be in the CGM and/or IGM with a temperature of $10^{5-7}$ K, which cannot be observed very easily \citep{bregman07,shull12,nicastro18,degraaff19}. The location of this gas, specifically what fraction is within versus outside the halo, is not clear. 

Because this paper focuses on simulations with bound outflows, the baryonic mass in the halo increases with time until the saturated state. From our simulations, the saturated state has $M_{h}$= (0.5-1.2)$\times 10^{10} M_\odot$ within 200 kpc radius (Fig. \ref{f:Mh_t_fountain_n0}); larger $M_h$ would violate the $L_X$ constraints (Fig. \ref{f:Lx_t_fountain_n0}). This mass is small compared to the mass of the missing baryons, which is around $7\times 10^{10}$ \msun. Therefore the majority of the missing mass, we argue, should be at $R>$ 200 kpc. This displacement of baryons relative to the DM halo can be due to much stronger SNe-driven outflows in the past which pushed gas in the halo out, and/or suppressed cosmic accretion from entering the halo. This effect is associated with the missing metals problem, which we discuss more quantitatively next.

\subsection{Missing metals problem}
\label{sec:missing_metals}

Besides the missing baryon problem, there is also the missing metals problem. Most metals do not reside in galaxies \citep{ferrara05,bouche07,peeples14,telford19}. \cite{peeples14} estimated that only 30\% of metals ever produced are in galaxies in a mass range of $10^{9-11.5}$ \msun. Like the missing baryon problem, the metals missing from galaxies are very likely in the CGM and/or IGM.

\begin{figure}
\begin{center}
\includegraphics[width=0.5\textwidth]{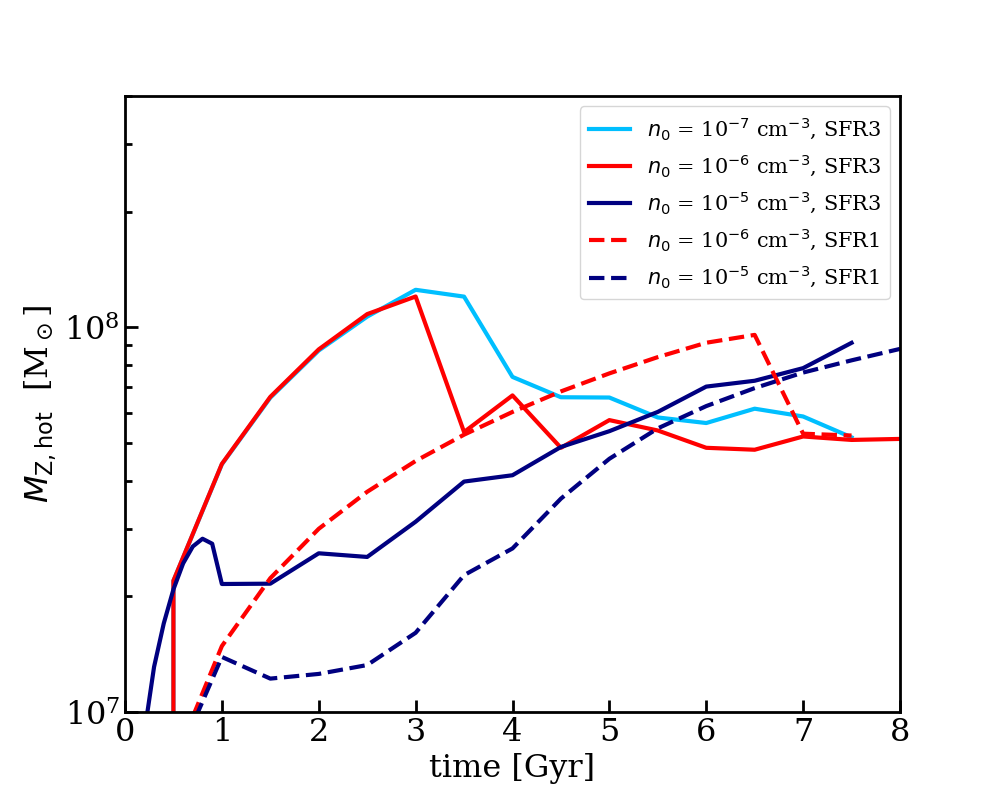}
\caption{Metal mass in the hot CGM. This does not include the metals in the pre-existing gas. For various models, $M_{\rm{Z,hot}}$ settles to (0.6-1.4)$\times 10^8$ M$_\odot$. }
\label{f:MZ_t_fountain_n0}
\end{center}
\end{figure}

We first show that the metals associated with the CGM in our simulations are small compared to the missing metals. Fig. \ref{f:MZ_t_fountain_n0} shows the total metal mass in the hot CGM, $M_{\rm{Z,hot}}$. Metals from the pre-existing gas in the halo are excluded. In other words, we only count metals from the hot atmosphere formed out of the SNe-driven outflows. The evolution of $M_{\rm{Z,hot}}$ is very similar to $M_h$. That is, the metal mass first increases and then reaches a plateau once the CGM is saturated. The final $M_{\rm{Z,hot}}$, (0.6-1.4) $\times 10^8$ M$_\odot$, depends little on $n_0$ or SFR. Observationally, the mass of metals in MW-mass galaxies is around $6\times10^8$ \msun, while the total metals ever produced are about $2\times 10^9$ \msun\ \citep{peeples14} (but note that there is at least a factor of 2 uncertainty related to the metal yield and the value of the solar metallicity). Our atmosphere thus does not contain a significant fraction of the missing metals. 

The maximum metal mass due to saturation in our simulations is for the late stage of the MW evolution, i.e., when the SF surface density is low and outflows are gravitationally bound by the halo. Metals that are still missing are likely carried away by galactic winds that are strong enough to leave the DM halo. The unbound winds can arise from earlier phases of galaxy formation, when the star formation and feedback are more intense \citep[e.g.][]{pettini01,bouche07,steidel10,rudie19,oppenheimer10,cen11,keller15}. 
Metal-enriched winds are also necessary to explain the observed metals in the IGM \citep[e.g.][]{songaila96,ellison00}.  In other words, we view the formation of the MW CGM as a two-step process: an early stage with unbound winds and a late stage with fountain flows, separated by the intensity of SF. With this picture in mind, the metals retained in the CGM only come from the later stage modeled here when outflows remain bound to the halo.

We will discuss the case of unbound winds in detail in a separate paper, but here we present a crude estimate of the total energy released by a MW-like galaxy in the form of SNe-driven hot outflows over cosmic time. From this estimation, the energy is sufficiently large to push metals out of the halo and suppress cosmic accretion, which will leave a low-density halo.  This low-density halo would then be the initial conditions for our current simulation.

The estimation utilizes a tight correlation, found from small-box simulations, between the energy and metal loading factor for the hot outflows, $\eta_{E,h}/\eta_{Z,h} \sim 0.5$ \citep{li20}.
This means that the missing metals from galaxies can be used to estimate the energy associated with these metals. For the MW, 30\% of metals residing in galaxies means that $\eta_{Z,h}\sim$ 0.7 over cosmic time, which gives $\eta_{E,h}\sim$ 0.35 from the above equation. The total amount of energy associated with these missing metals is then
\begin{equation}
    E_{\rm{out}}  =  2.1 \times 10^{59}\ \rm{erg} 
     \left( \frac{\eta_E}{0.35} \right)
    \left( \frac{N_{SN}}{ 10^9} \right)
    \left(\frac{E_{SN}}{10^{51} \rm{erg}}\right) \left(\frac{\alpha}{0.6}\right),
\end{equation}
where $N_{SN}$ is the total number of SNe exploded over the lifetime of the galaxy, and $\alpha$ is the fraction of hot outflows that have \vout$>$ $v_{\rm esc}$ over cosmic time. (If \vout$<$ $v_{\rm esc}$, the amount of metals retained in the CGM is not very large, as is the case in this paper.) The value of $\alpha$ is uncertain. Since $\phi \sim 2 \times 10^{15}$ erg g$^{-1}$ for MW halo (from the galaxy to $R\sim$250 kpc), this means that the $M \sim E_{\rm{out}}/\phi \sim 5.5\times 10^{10}$ \msun\ mass can be expelled out of the halo. The available energy from outflows is sufficient to keep the majority of the missing baryons, together with the missing metals, outside of the DM halo. Note that this is a conservative estimate since the galaxy potential is shallower in the past, and not all baryons fall to the center of the potential.

\subsection{Cosmic inflows}

Cosmic inflows are not modeled explicitly in our simulations. Gas can come into the halo in two forms: hot, nearly spherical accretion \citep{rees77,silk77}, and cool, filamentary inflows along cosmic webs \citep[e.g.][]{birnboim03,keres05}. The former is more important in massive halos and at low redshifts, which is the case for the MW in the recent past.
In our simulations, the pre-existing gas exerts weight on the outflows, mimicking the effects of the hot-mode accretion to some extent. Future work will include modeling this hot mode accretion explicitly \citep[e.g.][]{fielding18}. That said, we want to point out that hot outflows, when powerful enough, can suppress the spherical inflows \citep[e.g.][]{somerville15,zinger20}. As discussed in the previous subsection, based on the current metal mass in the halo, at early times outflows were so strong that they drove gas into the IGM.  
The low $n_0$ required in our simulations to reproduce the observed CGM properties, especially $L_X$ and N (O VI), does imply that suppression of inflows happened in the past (unless significant mass is ejected in the form of low volume-filling cold phase). 

We thus conjecture that the ``assembly" of the hot halo around the MW is a two-step process: first the strong hot outflows have left a halo with a relatively low mean density, $\lesssim 10^{-5}$ cm$^{-3}$; for the more recent past, less powerful hot outflows have led to the formation of a fountain atmosphere.  Note that the strong outflows with \vout$>v_{\rm{esc}}$ do not necessarily need a very large SFR; a mild SFR in the inner part of the galaxy where the gravitational field is large can also lead to a strong outflow \citep{li17a,armillotta19}.  The switch could have happened a few Gyr ago, as the thin disk of the MW is formed from inside out \citep[e.g.][]{bovy16,goddard17,telford19}. 

On the other hand, hot outflows cannot suppress the cold-mode accretion as efficiently. 
Observationally, cool CGM exists at both inner and outer part of the halo, and has a wide range of metallicities \citep[e.g.][]{lehner13,prochaska17}. The latter may point to different formation channels of cool CGM. In our simulations, cool gas only forms in the inner halo and has a large metallicity; it does not form at larger radii because the cooling time there is very long. 
The observed cool CGM that has a low metallicity or exists at large radii can be related to the cold-mode accretion and/or the stripping of satellite galaxies \citep[e.g.][]{hummels19,afruni19,hafen19}. Indeed, the similarity of cool CGM around star-forming and quiescent galaxies at large radii \citep[e.g.][]{chen10,tumlinson13,lan14} may indeed point to that this cool gas relates to the cosmic accretion, rather than feedback.

\subsection{Comparison to other work}

There are two recent analytic works by \cite{faerman20} and \cite{voit19} which have radially decreasing temperature profiles, broadly consistent with our simulation results. Both models reproduce the observed O VI and $L_X$ of the MW. And their predicted Ne VIII and Mg X are very similar to ours (except at large radii where photoionization would make a difference). Here we comment on the comparisons between their models and our simulation. One difference is that while they assume a steady-state CGM, our simulations have time evolution. One key assumption \cite{faerman20} made is that the atmosphere is isentropic. In our dynamic model it is naturally the case within \rmax\ before saturation; after saturation, the CGM starts to deviate from isentropic. The evolution of N (O VI) reflects this transition.
Their model is tuned so that N (O VII) and N (O VIII) match the observations, whereas our simulations self-consistently produce the observed N (O VII) \& (N O VIII). \cite{voit19} presents a precipitation-limited model of the CGM by assuming  $t_{\rm{cool}}/t_{\rm{free-fall}} =10-20$ and hydrostatic equilibrium for all radii. In our simulations, the precipitation-limit is reached, but only in the inner halo $R\lesssim$ 50 kpc. Beyond that, the ratio $t_{\rm{cool}}/t_{\rm{free-fall}}$ is larger than 20 throughout, i.e., the outer halo never reaches the precipitation limit. Similar results are seen in simulations by \cite{fielding17}. 

In terms of the hot CGM mass, our results are very similar at $R\lesssim$ 60 kpc, which is (2-4) $\times 10^9 M_\odot$ (this is also well-constrained from observations, see Section \ref{sec:univ_den}). The difference is more obvious at large radii: our mass within 200 kpc is 0.5-1.2$\times10^{10}$ $M_\odot$, compared to their mass of $3\times10^{10}$ $M_\odot$. This is mainly because the density in our simulation drops quickly with radius due to the confinement of outflows by gravity. By assuming non-thermal pressure support \citep{faerman20}, or a precipitation limit \citep{voit19}, their density profile drops slower than ours at $R\gtrsim$ 60 kpc.

Cosmological simulations have the advantage of modeling the CGM with cosmic inflows and satellite galaxies orbiting within the halo \citep[e.g.][]{stinson12,ford14,oppenheimer18,grand19}. These processes can contribute to the cool CGM phases, especially at large radii, which we do not cover. That said, recent cosmological simulations with enhanced numerical resolution in the CGM region show that the properties of the cool CGM do not converge \citep{peeples19,voort19,hummels19,suresh19}. In addition, cosmological simulations use \textit{ad hoc} models for feedback, which likely do not generate outflows with the same properties as ours, thus the thermal, dynamical and chemical states of the CGM/IGM would be different, since CGM is sensitive to the feedback schemes and strengths \citep[e.g.][]{nelson18,davies20}. A comparison study between idealized simulations and cosmological ones is underway (Fielding et al., in prep). 
Since the CGM is where cosmic inflows interact with galactic outflows, a predictive model of the CGM needs both a robust feedback model and a cosmological context.  Future endeavors should combine realistic small-scale feedback physics together with a cosmological \textit{ab initio} condition. Recent zoom-in simulations like FIRE are approaching this goal for small galaxies \citep{hopkins13b,muratov17}.

\section{Conclusions}

From small-box simulations, the hot phase of SNe-driven outflows are found to be the dominating phase, which carry the majority of outflow energy and metals, and can travel much further than the cooler phases. 
In this paper, we investigate how the hot outflows emerging from a MW-like galaxy (in terms of gravitational potential and star formation) evolve on large scales. The loading factors of hot outflows are taken from small-box simulations for the current SF surface density of the MW. From the loading factors, the specific energy of the hot outflows is not sufficient for them to escape from the halo. Indeed, a large-scale fountain is formed. Our main findings are the following:

\begin{enumerate}
\item The hot SNe-driven outflows form a metal-enriched, warm-hot atmosphere in the halo, with fountain motions. 
The maximum radius the outflows reach, \rmax, can be larger than the radius from a simple energy argument \rout\ (defined in Eq. \ref{eq:rout}), indicating outflows from different SF episodes do not evolve in isolation.

\item  For a given set of loading factors, \rmax\ is smaller when there is more pre-existing gas; however, \rmax\ does not depend on the rate at which the outflows are injected (Fig. \ref{f:rmax_diff_n0_SFR}). 

\item As more mass accumulates in the halo, the inner CGM is saturated, i.e., the mass of hot gas reaches a maximum. This steady-state is maintained because cool clumps condense out of the hot atmosphere and fall toward the galaxy ballistically (Fig. \ref{f:vr_r_hot_cool}). This is a natural way of forming the highly-enriched ``high-velocity" clouds.

\item After the saturation is reached, the condensation of cool gas balances the hot outflow injection (Fig. \ref{f:Mh_t_fountain_n0}).

\item The balance leads to a universal density profile up to \rmax, which has a break at \rout\ (Fig. \ref{f:n_r_t_diff_SFR}). 
This universal density profile does not depend on the star formation rate.

\item The hot CGM has a radially-decreasing temperature profile, due to the expansion of hot outflows (Fig. \ref{f:entropy_T_r_t_1e-7_SFR3}). Together with the density profile, several important CGM observables are naturally reproduced, including X-ray luminosity and column densities of O VI, O VII, O VIII, assuming CIE (Fig. \ref{f:Lx_t_fountain_n0}, \ref{f:N_O678}, \ref{f:N_OVI_t}). 

\item The collisionally-ionized O VI is located in a shell at 100-150 kpc, which cools inefficiently. The formation of an O VI-shell requires a small mean density of pre-existing gas $n_0\lesssim10^{-6}$ cm$^{-3}$ which allows \rmax$\gtrsim$ 100 kpc.

\item Saturation determines the maximum mass of baryon and metals in the hot atmosphere, which are about $(0.5-1.2)\times 10^{10}\ M_\odot$ and $(0.6-1.4) \times 10^8\ M_\odot$, respectively. These are not significant amounts compared to the ``missing baryons" and ``missing metals". We conjecture that the missing metals reside at even larger radii and were ejected from an unbound galactic wind at earlier epochs of galaxy formation.

\end{enumerate}

The CGM is known to be complex and likely many processes are ongoing, but we hope to better understand it by starting with fewer assumptions and by isolating the number of processes involved. 
Our simulations have simple inputs: hot outflows and hot pre-existing halo gas. The resultant CGM is multiphase: a hot atmosphere ($> 10^6$ K), a warm-hot shell ($\sim 10^5$ K) from the expansion of hot outflows, and cool clumps ($\sim 10^4$ K) precipitating due to the condensation of the hot atmosphere. Despite its simpleness, the CGM reproduces many aspects of observations. We also make predictions for intermediate and high ions such as Ne VIII, Ne IX and Mg X, which can be compared against future observations. Finally, feedback from supermassive black holes can also impact the CGM, as evidenced by the Fermi bubble \citep{su10}. It will be interesting to investigate in the future how black holes and SNe together affect the CGM.

\section*{Acknowledgement}
We thank the referee for the comments for this paper. ML thanks the insightful discussions with many colleagues, including Chris McKee, Crystal Martin, Cameron Hummels, Daniel Wang, Eve Ostriker, Yuan Li, Greg Bryan, Mark Voit, Taotao Fang, Mary Putman, Josh Peek, Jason Tumlinson, Joel Bregman, Yakov Faerman, Smita Mathur, Drummond Fielding, Filippo Fraternali, and Amiel Sternberg. Computations were performed using the publicly-available Enzo code, which is the product of a collaborative effort of many independent scientists from numerous institutions around the world. Their commitment to open science has helped make this work possible. Data analysis and visualization are partly done using the \textsf{yt} project \citep{turk11}. The simulations are performed on the Rusty cluster of the Simons Foundation. We thank the Scientific Computing Core of the Simons Foundation for their technical support.

\appendix
\restartappendixnumbering

\section{Outflow fluxes}
\label{sec:flux}

\begin{figure}
\begin{center}
\includegraphics[width=0.48\textwidth]{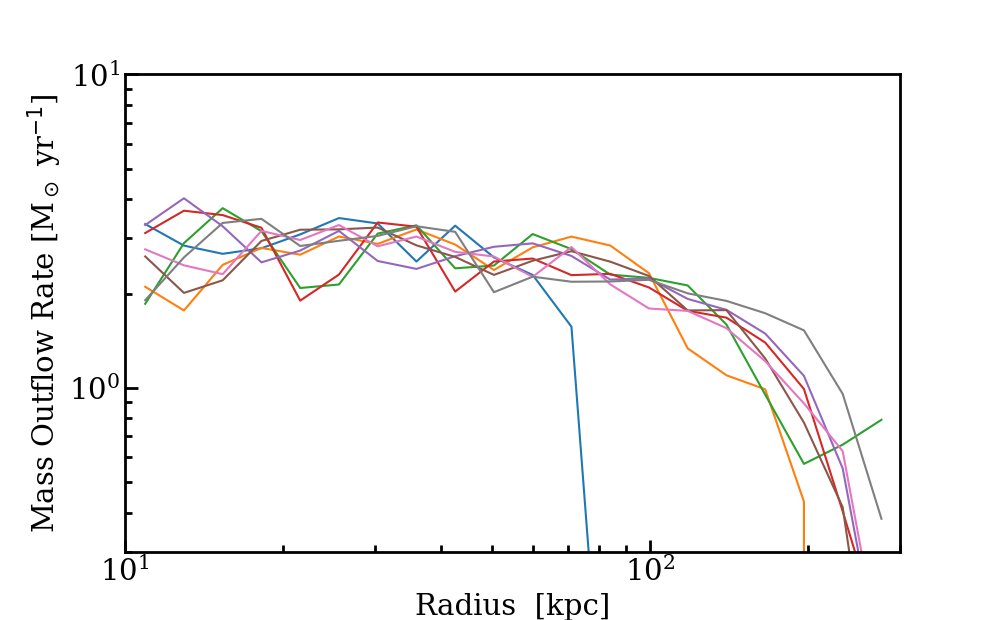}
\includegraphics[width=0.48\textwidth]{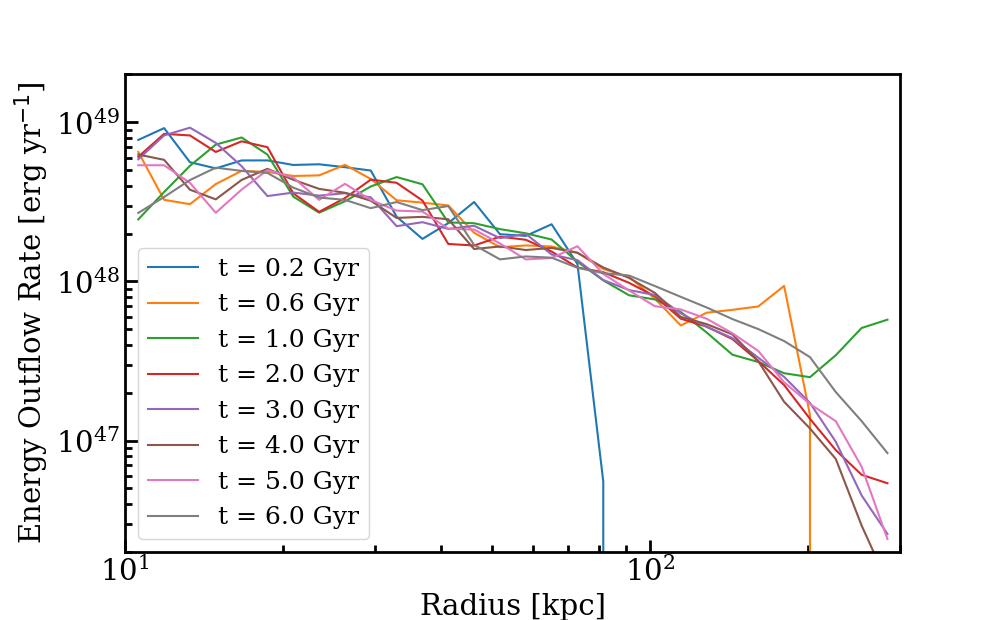}
\caption{Mass outflow rate (left panel) and total energy outflow rate (right) as a function of radius, for a test run 1e-6SFR3 with cooling turned off. This is to show that the injection rates of mass and energy are as expected, which are 3 $M_\odot$ yr$^{-1}$ and 9$\times 10^{48}$ erg yr$^{-1}$, respectively, at the inner boundary.  }
\label{f:fluxes}
\end{center}
\end{figure}

To check that the outflow rates agree with our inputs from the small box simulations \citep{li17a}, we run a test of 1e-6SFR3 with cooling turned off, and show the outflow rates of mass and energy (including kinetic and thermal forms) as a function of radius in Fig. \ref{f:fluxes}. The expected rate at the inner boundary are 3 $M_\odot$ yr$^{-1}$ and 9$\times 10^{48}$ erg yr$^{-1}$, respectively, which are taken from the small box simulations. The simulation shows good agreement. Time variations are expected since injections are discrete events. The mass outflow rate is flat and then drops at the turn-around radius of 130 kpc. The energy outflow rate declines gradually with radius as a fraction of gas energy is converted into gravitational energy. 

\section{X-ray luminosity}
\label{sec:Lx_R}

\begin{figure}
\begin{center}
\includegraphics[width=0.48\textwidth]{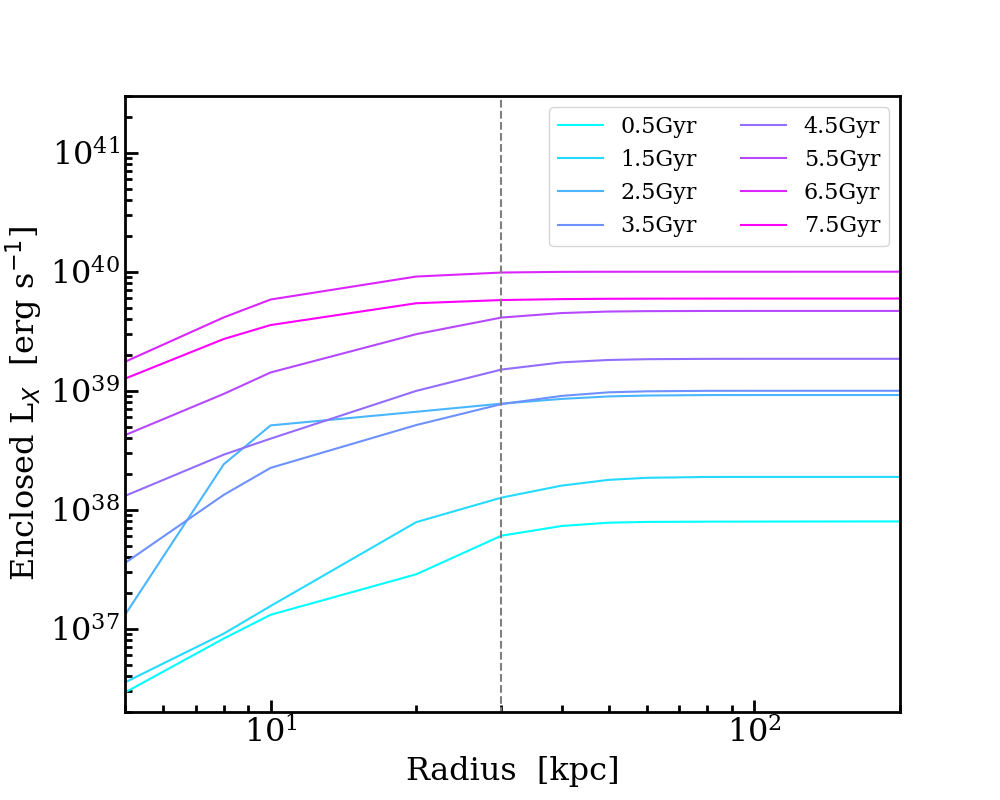}
\includegraphics[width=0.48\textwidth]{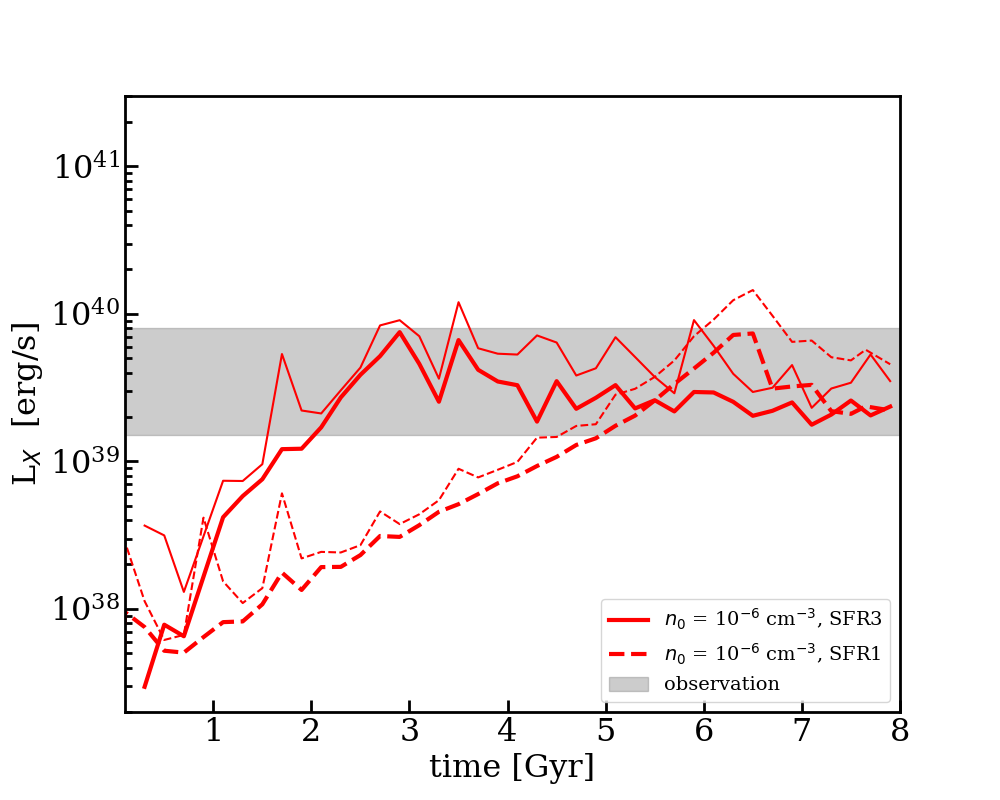}
\caption{Left panel: spherically enclosed X-ray luminosity as a function of radius for 1e-6SFR1. Most emissions come from the inner 30 kpc (dashed line). Right panel: X-ray luminosity at 0$<R<$ 30 kpc (thin line) and at 10$<R<$ 30 kpc (thick line). }
\label{f:Lx_R}
\end{center}
\end{figure}

In Figure \ref{f:Lx_R} we show the radial dependence of the X-ray luminosity.  This illustrates why we choose 10$<R<$ 30 kpc as the region to calculate the X-ray luminosity $L_X$ in Figure \ref{f:Lx_t_fountain_n0}. The left panel shows that most X-ray emission comes from  $R<$ 30 kpc. The right panel shows $L_X$(0$<R<$ 30 kpc) as thin lines and $L_X$(10$<R<$ 30 kpc) as thick lines. $L_X$(0$<R<$ 30 kpc) is biased toward higher luminosity because of the temporal sampling of data outputs: the time interval of data output (every 10 Myr) is close to that of the outflow injection (every 9.9 Myr), therefore more snapshots are taken immediately after the outflow injection, when $L_X$ is higher compared to the time-averaged mean, which we deem as the ``true value". In contrast, $L_X$(10$<R<$ 30 kpc) is less than the true value due to the exclusion of the inner region. In other words, the true time-averaged $L_X$ should be in between these two lines. But since the two curves are closer than the observation scatter (grey region), we use the latter as a proxy in the paper given its smoother feature. 

\section{Resolution test}
\label{sec:res}

\begin{figure}
\begin{center}
\includegraphics[width=0.50\textwidth]{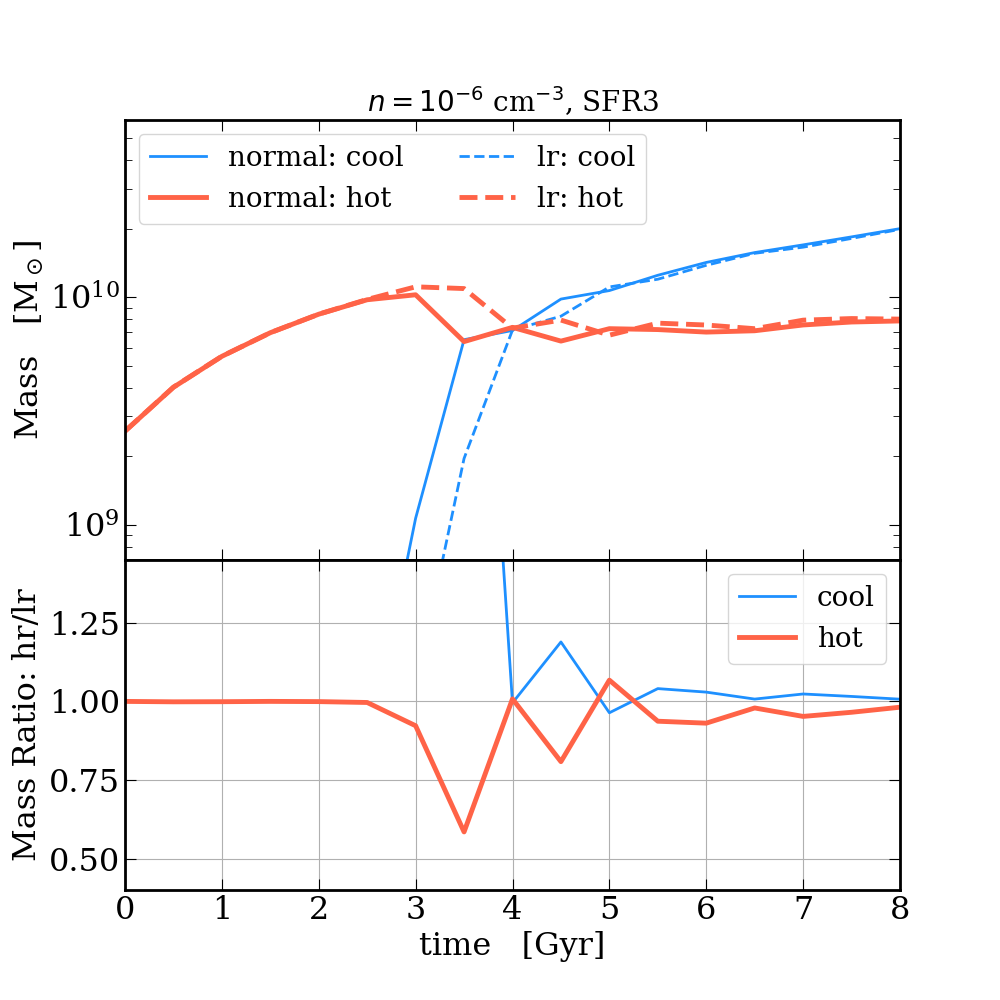}
\caption{Resolution check of mass of hot and cool gas in the simulation box as a function of time. The run for comparison is 1e-6SFR3. The top panel shows the mass, where the solid lines represent the fiducial resolution, and the dashed lines are for the run with a resolution coarser by a factor of 2. The bottom panel shows the mass ratio between the fiducial run and the low-resolution one . The fiducial run has a slightly earlier onset of cooling, but the overall agreement is good after that.}
\label{f:m_res}
\end{center}
\end{figure}

We verify that our results are robust for the run 1e-6SFR3 because it shows strong agreement with observations.
Fig. \ref{f:m_res} shows the resolution comparison for the mass of the cool and hot gas in the simulation box. The general agreement is very good. The largest difference is when the cool gas begins to form, around 3-4.5 Gyr. The normal-resolution run has an earlier onset of cooling. This is expected since the the higher resolution run, with eight times more cells, samples a broader distribution of cooling time. After that the agreement is good. Other properties of the CGM, such as \rmax\, the radial profiles of density, temperature, entropy, and observables such as the column densities of different ions, show little change with resolution.

\begin{figure}
\begin{center}
\includegraphics[width=1\textwidth]{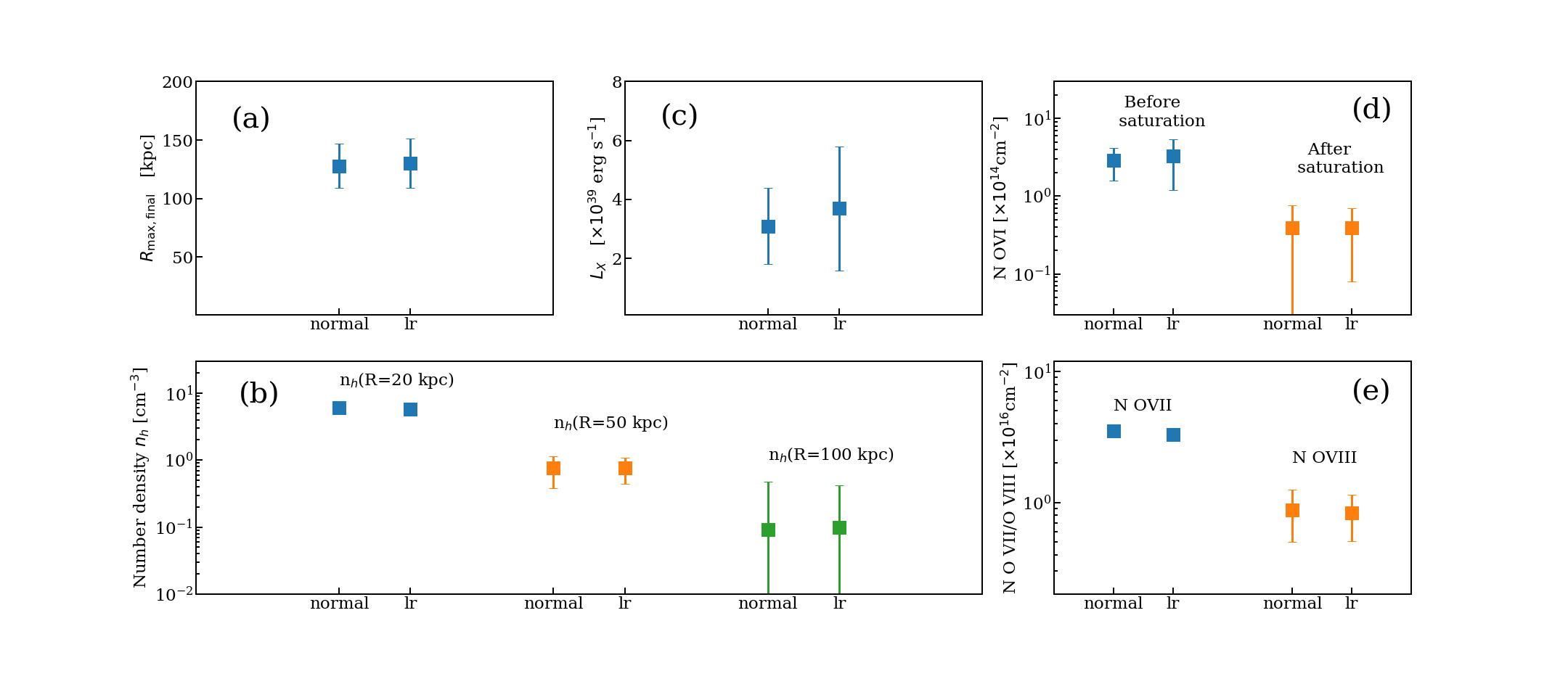}
\caption{Resolution checks for 1e-6SFR3. Same as for the previous figure, ``normal" is for the fiducial run and "lr" indicates the run with a resolution 2 times coarser. Panel (a) $R_{\rm max}$, (b) mean number density of hot gas at a few radii, (c) X-ray luminosity L$_X$, (d) column density of O VI from sight lines with impact parameters $50<d<$100 kpc, (e) column density of O VII and O VIII from sight lines going through the center of the box.  The quantities shown in panels (a)-(c),(e) are averaged over $t=3-8$ Gyr, after the saturation state of the CGM is reached and the system enters a quasi-steady state. For N (O VI), we show the comparison for the two stages: before saturation (2-3 Gyr), and after saturation (4-5 Gyr). The error bars indicate the standard deviation of the temporal variations for (b) and (c), while showing the temporal and sight-line variations for (a), (d), (e). All quantities show little changes with respect to the resolution.  }
\label{f:res_5panel}
\end{center}
\end{figure}

We show other quantities for the resolution check in Fig. \ref{f:res_5panel}. The quantities, except N (O VI) in panel (e) are calculated by averaging over $t=3-8$ Gyr, after the CGM reaches the saturated state and these quantities show quasi-steady values. As we show in Fig. \ref{f:N_OVI_t}, N (O VI)  has two evolutionary states, with a higher value before the reverse shocks heat the outer radii than after. We therefore check the resolution effect on N (O VI) at these two states, respectively. The error bars show the standard deviation of the temporal variations for $n_h$ and $L_X$, while showing that of the temporal and sight-line variations for \rmax, N (O VI), N (O VII), and N (O VIII). From the figure, a factor of 2 change in resolution leads to little difference on these quantities. This is not surprising since these quantities come from well-resolved gas structure in the simulation.

\vspace{0.2in}

%\bibliography{master_bib}
%\bibliographystyle{aasjournal}

\end{CJK*}
\end{document}